\documentclass[11pt]{article}
\pdfoutput=1
\usepackage{jnpub,amsmath,amssymb,hyperref}

\usepackage{graphicx}
\usepackage{todonotes}
\usepackage[mathscr]{eucal}
\usepackage{amsmath}
\usepackage{color}
\usepackage{epsfig}
\usepackage{epstopdf}
\usepackage[utf8]{inputenc}

\newcommand{\gsimm}{\raise.3ex\hbox{$>$\kern-.75em\lower1ex\hbox{$\sim$}}}
\newcommand{\lsimm}{\raise.3ex\hbox{$<$\kern-.75em\lower1ex\hbox{$\sim$}}}

\newcommand{\comment}[1]{}

\newcommand{\St}{{St\"uckelberg}}

\newcommand{\be}{\begin{equation}}
\newcommand{\ee}{\end{equation}}
\newcommand{\ba}{\begin{eqnarray}}
\newcommand{\ea}{\end{eqnarray}}

\newcommand{\bea}{\begin{eqnarray*}}
\newcommand{\eea}{\end{eqnarray*}}

\def\a{\alpha}

\begin{document}
\title{Dark Energy and Doubly Coupled Bigravity}

\author[a]{Philippe Brax,}
\author[b]{Anne-Christine Davis,}
\author[c]{Johannes Noller}

\affiliation[a]{Institut de Physique Th\'{e}orique, Universit\'e Paris-Saclay, CEA, CNRS, F-91191 Gif/Yvette Cedex, France}
\affiliation[b]{DAMTP, Centre for Mathematical Sciences, University of Cambridge, CB3 0WA, UK}
\affiliation[c]{Astrophysics, University of Oxford, DWB, Keble Road, Oxford, OX1 3RH, UK}

\emailAdd{philippe.brax@cea.fr}
\emailAdd{A.C.Davis@damtp.cam.ac.uk}
\emailAdd{noller@physics.ox.ac.uk}

\date{today}
\abstract{We analyse the late time cosmology and the gravitational properties of doubly coupled bigravity {in the {constrained} vielbein formalism} {(equivalent to the metric formalism)} when the mass of the massive graviton is of the order of the present Hubble rate. We focus on one of the two branches of background cosmology where the ratio between the scale factors of the two metrics is algebraically determined. We find that the late time physics depends on the mass of the graviton, which dictates the future asymptotic cosmological constant. The Universe evolves from  a matter dominated epoch to a dark energy dominated era where the equation of state of dark energy can always be made close to -1 now by appropriately tuning the graviton mass. We also analyse the perturbative spectrum of the theory in the quasi-static approximation, well below the strong coupling scale where no instability is present, and we show that there are five scalar degrees of freedom, two vectors and two gravitons. In Minkowski space, where the four Newtonian potentials vanish, the theory {manifestly} reduces to one massive and one massless graviton.  In a cosmological FRW background for both metrics, four of the five scalars are Newtonian potentials which lead to a modification of gravity on large scales. The fifth one gives rise to a ghost which decouples from pressure-less matter in the quasi-static approximation. In this scalar sector, gravity is  modified with effects on both the growth of structure and the lensing potential. In particular, we find that the $\Sigma$ parameter governing the Poisson equation of the weak lensing potential can differ from one in the recent past of the Universe. Overall, the nature of the modification of gravity at low energy, which reveals itself in the growth of structure and the lensing potential, is intrinsically dependent on the couplings to matter and  the potential term of the vielbeins.  We also find that the time variation of Newton's constant  in the Jordan frame can easily satisfy the bound from solar system tests of gravity. Finally we show that the two gravitons present in the spectrum have a non-trivial mass matrix whose origin follows from the potential term of bigravity. This mixing leads to gravitational birefringence.}



\maketitle

\setcounter{tocdepth}{2}

\section{Introduction}
The late time acceleration of the expansion of the Universe could be linked to a modification of gravity on large scales \cite{Koyama:2015vza}. In fact, dark energy, i.e. the presence of a new form of matter leading to the acceleration of the expansion \cite{Copeland:2006wr},  and modified gravity, i.e. a change in the gravitational dynamics compared to General Relativity (GR),  are not mutually exclusive \cite{Joyce:2014kja} and many models lead to both phenomena. This is certainly true of all the screened models of modified gravity \cite{Khoury:2010xi} such as $f(R)$ theories \cite{Hu:2007nk} in the large curvature limit, K-mouflage \cite{Babichev:2009ee} or Galileons \cite{Nicolis:2008in}, which display either the chameleon \cite{Khoury:2003rn}, K-mouflage or Vainshtein \cite{Vainshtein:1972sx} screening mechanisms. In all these models, a scalar field is singled out and its role is to induce changes to both the background cosmology and the growth of structure compared to the $\Lambda$-CDM template. Sometimes, as for $f(R)$ models, the difference only really shows up at the perturbative level \cite{Brax:2012gr}. Other times, for K-mouflage \cite{Brax:2014wla} and Galileons \cite{Appleby:2011aa}, both the background and perturbative properties of the models differ from $\Lambda$-CDM.

Another and maybe more fundamental approach has been pursued in the last few years and consists in analysing the behaviour of consistent field theories going beyond GR. A particularly relevant example is ghost-free massive gravity \cite{deRham:2010ik, deRham:2010kj, Hassan:2011vm}, a ``bimetric'' theory which involves a single dynamic metric and another passive one. In ghost-free massive bigravity  \cite{Hassan:2011hr, Hassan:2011zd}, the second metric is promoted to a dynamical variable while matter minimally couples to one of the two metrics only.
Consistent extensions\footnote{To be explicit, we take ``consistent'' to mean that the theory has a low-energy limit with non-trivial {(non-linear and in our case typically irrelevant)} interactions that is ghost-free. Whether requiring ghost-freedom beyond this limit/energy scale is a physically meaningful criterion depends on whether one is willing to trust a theory beyond the regime where perturbative unitarity is lost.} of these bigravity theories with non-derivative matter couplings that involve both metrics have been found in \cite{deRham:2014naa,Noller:2014sta,Melville:2015dba}. \footnote{For a discussion of extensions involving derivative matter couplings see \cite{Heisenberg:2015wja,Gao:2016vtt}.}

All these approaches are frequently plagued with instabilities and/or inconsistencies and incompatibilities with observations, both at the background and perturbative levels. For massive gravity, it has proved impossible to find consistent and flat FRW background solutions \cite{D'Amico:2011jj} (although solutions which approximate such FRW backgrounds to great accuracy exist). In the singly coupled bigravity case, this obstacle can be overcome  \cite{Volkov:2011an,vonStrauss:2011mq,Comelli:2011zm}, while perturbations in the scalar, vector  and tensor sectors can show power law or exponential instabilities \cite{Comelli:2012db,Lagos:2014lca,Cusin:2014psa,Konnig:2014xva,Amendola:2015tua,Johnson:2015tfa,Konnig:2015lfa,Cusin:2015pya}.
Finally, in doubly coupled bigravity models as we are considering here, there are two branches of viable background solutions \cite{Enander:2014xga,Lagos:2015sya} and the perturbative properties of these models have been partially explored in \cite{Comelli:2015pua, Gumrukcuoglu:2015nua}, with results suggesting that {they} might be improved with respect to the singly coupled case. Note that the couplings of \cite{deRham:2014naa,Noller:2014sta}, upon freezing one of the metrics/vielbeins, also straightforwardly give rise to new massive gravity (i.e. non-bigravity) couplings, whose features we will discuss further separately in \cite{us}.

In this paper, we will use the constrained vielbein formulation of bigravity doubly coupled to matter. {The constraint ensures that our theory here is equivalent to the metric formulation, whereas in general the unconstrained vielbein \cite{Noller:2014sta} and metric \cite{deRham:2014naa} ``formulations'' are not equivalent} \cite{Noller:2014sta,Hinterbichler:2015yaa,deRham:2015cha,Melville:2015dba}. Here we will therefore explicitly enforce the symmetric vielbein condition \cite{Deser:1974zzd} {from the start}, which does ensure that the two formulations of bigravity are equivalent \cite{Deffayet:2012zc} (and which is in fact dynamically enforced, without the need for an explicit constraint, in the low-energy/decoupling limit of these theories \cite{deRham:2015cha,Melville:2015dba}). {Note that, in general and beyond the decoupling limit, when not working with constrained vielbeins from the start, it is known} {that in the doubly coupled case and in the vielbein formulation, the symmetric condition cannot always be imposed consistently afterwards \cite{Hinterbichler:2015yaa}. In this paper we will {therefore use} constrained vielbeins {satisfying the symmetric condition} when we study the dynamics of the theory, and couple matter to the Jordan metric built out of a linear combination of constrained vielbeins. }

We will  be mostly preoccupied with late time properties in the late radiation, matter and dark energy eras at the background and scalar perturbation levels. {Focusing mostly on the late-time properties of the theory is partially motivated by the very low strong coupling scale $\Lambda_3= (m_{\rm Pl} m^2)^{1/3}$ of the model. Above this scale loop corrections cannot be ignored and blindly trusting the tree-level calculation becomes a significant leap of faith\footnote{Note that this scale depends on the background and it has recently been suggested that, for (approximately Lorentz-invariant) backgrounds different from the precisely Lorentz-invariant Minkowski background considered here, the strong coupling scale could potentially be raised from $\Lambda_3$ up to $\Lambda=(mm_{\rm Pl})^{1/2}$ where $m\ll m_{\rm Pl}$ \cite{deRham:2016plk}.}. Therefore the low-energy phenomenology of the theory in a sense provides the most conservative and robust observational test bed for the theory. In other words, if there is at least some regime where the theory is in fact realised in nature, it has to be this one, whereas at higher energies the precise predictions of the theory should rely heavily on its UV completion. As such, investigating our theory in the late universe/low energy regime is of intrinsic interest.}

We consider the cosmology and gravitational properties of doubly coupled bigravity below the strong coupling scale $\Lambda_3$. When the graviton mass of order $m$ is taken to be similar to the Hubble rate now $H_0\sim 10^{-42}$ GeV, the strong coupling scale is $\Lambda_3 \sim 10^{-22}$ GeV. This implies that we only consider scales larger than $\Lambda_3^{-1}\sim 1000$ km, which allows one to study gravitational properties of planetary orbits in the solar system for instance.  Cosmologically we are only describing the eras for which $H\lesssim \Lambda_3$ which corresponds to redshifts $z\lesssim 10^{11}$, i.e. from the time of Big Bang Nucleosynthesis to now. In practice we will restrict ourselves to the study of the late radiation,  matter and dark energy eras. Numerically we will set the initial conditions at the matter-radiation equality. At the background cosmological level
we retrieve the known result that two branches of solutions exist \cite{Enander:2014xga,Lagos:2015sya,Comelli:2015pua,Gumrukcuoglu:2015nua} in the presence of a perfect fluid and we focus on the branch where the two scale factors and the lapse functions are directly related. In this formulation, the matter-radiation eras are followed by a dark energy epoch whose characteristics depend on the graviton mass and the coefficients of the vielbein's potential term. In these eras and in the Jordan frame, the scalar perturbations of the metric can be described by two Poisson equations for the Newtonian potentials of the Jordan frame metric. After normalising Newton's constant to local gravitational tests -- which can be easily satisfied for distances much smaller than the graviton's Compton wavelength, i.e. standard gravity is retrieved at short distance with no need for a screening mechanism, when the two couplings to matter are present \footnote{Were it that one coupling should disappear, i.e. in the limit where our double coupling reduces back to the minimally singly coupled case, this result would not hold.} -- we find that cosmological perturbations deviate from $\Lambda$-CDM provided the ratio of the two lapse functions differs in the  matter era and the dark energy one.  As a result, the background evolution,  the growth of structure and the lensing properties of the models deviate from $\Lambda$-CDM at late times even though one can tune the graviton mass in order to fix the dark energy scale today.

We also come back to the general issue of cosmological perturbations in bigravity. For this we analyse the scalar, vector and tensor perturbations {when imposing the symmetric conditions}. We find that there are only 14 physical degrees of freedom: 6 scalars, 2 divergence-less vectors and 2 traceless transverse tensors.  The six scalar modes comprise four Newtonian potentials and two extra scalars. In the quasi-static approximation, which befits late-time cosmology and local physics in the presence of static sources, the number of physical scalars reduces to five comprising four Newtonian potentials. The fifth scalar has a higher order action in derivatives and can be described by two {second-order} scalar fields, one of them being a ghost.  In a FRW background, the four Newtonian potentials lead to late time modified gravity, which we have already described. We also find that the two vector fields do not receive potential terms. One decays at late time whilst the other one decouples from matter and can be set to be vanishing in the quasi-static approximation. In a Minkowski background, the two gravitons {manifestly} become one massive and one massless ones. In an FRW background, the two gravitons mix and give rive to gravitational birefringence.

The paper is arranged as follows. In section 2, we derive the Einstein equations and analyse their solutions in the FRW case. We retrieve the existence of two branches from the compatibility of the Friedmann equations and the Raychaudhuri equations. In section 3, we consider scalar perturbations and find that in the quasi-static approximation they reduce to four Newtonian potentials. We then analyse the Poisson equations for the Newtonian potentials in the Jordan frame and define the parameters $\eta,\mu$ and $\Sigma$ which characterise the deviations of cosmological perturbations from GR. We also analyse the vector and tensor perturbations in the quasi-static approximation. In section 5, we consider the background cosmology in the matter-radiation and dark energy eras and the instabilities in the radiation era. In section 5, we focus on the local dynamics in Minkowski space around overdensities with small Newtonian potentials. We find that GR is retrieved in this limit and this allows us to identify the local Newton constant. In section 6, we explore two typical models where the coupling constants differ (model I) or the coefficients of the vielbein potential are different (model II) and we solve the background equations of motion in this case. This allows us to discuss the deviation of the Hubble rate from its $\Lambda$-CDM counterpart, and the evolution of the parameters $\eta,\mu$ and $\Sigma$ with the redshift. In particular we find that gravity is not modified deep in the matter era and in the future dark energy era. As such, when the quasi-static approximation applies, gravity is only altered transiently between the matter and dark energy eras. Finally we have added an appendix on cosmological perturbations.

\section{Bigravity}
\subsection{Einstein's equations}

We consider massive bigravity models coupled to matter in the {constrained} vielbein formalism for energy scales below the strong coupling limit $\Lambda_3$ (note that this is different from the Vainshtein scale). This will allow us to study gravitational properties of planetary orbits in the solar system and cosmology after Big Bang Nucleosynthesis\footnote{ Deep inside the solar system on scales $r\lesssim 1000$ km our results would certainly need to be altered.}. This uses two {constrained} vielbeins $e_{1\mu}^a$ and $e_{2\mu}^a$ which couple to matter with couplings $\beta_{1,2}$ respectively. Although we will use the two vielbeins throughout the paper, this formulation of bigravity is equivalent to the metric one where the two metrics built from the two vielbeins are taken as the fundamental degrees of freedom. The equivalence between the two presentations is guaranteed by the symmetric condition \eqref{sym1}.

 The action comprises three very distinct parts. The first one is simply the Einstein-Hilbert terms
\be
S_G= \int d^4x\ e_1 \frac{R_1}{16\pi G_N} + \int d^4 x\ e_2 \frac{R_2}{16\pi G_N}
\ee
where $R_{1,2}$ are the Ricci scalars built from the respective vielbeins, and $e_{1,2}$ are the determinants of the vielbeins viewed as $4\times 4 $ matrices.
{Matter  fields $\psi_i$ are (minimally) coupled to the Jordan metric built from the local frame \cite{Noller:2014sta}
\be
e^a_\mu = \beta_1 e_{1\mu}^a+\beta_2 e_{1\mu}^a
\ee
where $a$ is a local Lorentz index and $\mu$ the global coordinate index associated with the one forms $e^a= e^a_\mu dx^\mu$.}
The matter action effectively consists of the coupling of the matter fields $\psi_i$ to the {Jordan metric $g_{\mu\nu}$
\be
S_m(\psi_i, g_{\mu\nu})
\ee
which is defined below.} { The matter action breaks the two copies of diffeomorphism and local Lorentz invariances which are preserved by the Einstein-Hilbert terms.}
{The individual vielbeins $e^a_{\alpha\mu}, \ \alpha=1,2,$ are constrained to satisfy the symmetric condition}
\be
e_{1\mu}^a e_{2\nu}^b \eta_{ab}=e_{1\nu}^a e_{2\mu}^b \eta_{ab}.
\label{sym1}
\ee
{ This  ensures the  equivalence with doubly coupled bigravity in the metric formulation. }
Massive bigravity also involves a potential term \cite{Hinterbichler:2012cn,Hassan:2011hr, Hassan:2011zd}
\be
S_V=\Lambda^4 \sum_{ijkl} m^{ijkl} \int d^4 x\  \epsilon_{abcd} \epsilon^{\mu\nu\rho\sigma} e^a_{i\mu} e^b_{j\nu}e^c_{k\rho} e^d_{l\sigma}
\ee
where
\be \Lambda^4= m^2 m^2_{\rm Pl}
\ee
and $m$ is related to the graviton mass while the dimensionless and fully symmetric tensor $m^{ijkl}$ involves five real coupling constants, which are all of order one as long as we adopt an effective field theory perspective. {Note that our $\Lambda$ corresponds to what is frequently denoted as $\Lambda_2$ in the literature. }
We have written the potential term as a function of the two vielbeins. The symmetric conditions (\ref{sym1}) allows one to rewrite $S_V$ as a function of the two metrics built from the two vielbeins.
{The two metrics are
\be
g^{\alpha}_{\mu\nu}= \eta_{ab} e^a_{\alpha\mu}e_{\alpha\nu}^{b}, \ \ \alpha=1,2
\ee
 and the corresponding Jordan metric
 \be
g_{\mu\nu}= \eta_{ab} e^a_\mu e^b_\nu
\ee
which is explicitly related to the $g^\alpha_{\mu\nu}$'s by
\be
g_{\mu\nu}= \beta_1^2 g^1_{\mu\nu} + \beta_1 \beta_2 Y_{\mu\nu} + \beta_2^2 g^2_{\mu\nu}
\ee
where we have defined the symmetric tensor
\be
Y_{\mu\nu}= \eta_{ab}( e^a_{1\mu} e^b_{2 \nu}+ e^a_{2\mu} e^b_{1 \nu}),
\ee
which can also be expressed as the square root of the ratio between the two metrics\cite{deRham:2014naa}. The overall result is that the full action can be expressed, albeit in a complex way, as a function of the two metrics
$g^{\alpha}_{\mu\nu}, \ \alpha=1,2$, solely.

{The Einstein equations can then be obtained by varying the action with respect to the two metrics {and can be written} formally as
\be
G^{1}_{\mu\nu} = 8\pi G_N ( T^{1}_{\mu\nu} + {\cal T}^{1}_{\mu\nu})
\ee
and
\be
G^{2}_{\mu\nu} = 8\pi G_N ( T^{2}_{\mu\nu} + {\cal T}^{2}_{\mu\nu})
\ee
where we have introduced the tensors
\be
T^{\alpha}_{\mu\nu}= -\frac{2}{e_\alpha} \frac{\delta S_m}{\delta g_\alpha^{\mu\nu}},\ \ {\cal T}^{\alpha}_{\mu\nu}= -\frac{2}{e_\alpha} \frac{\delta S_V}{\delta g_\alpha^{\mu\nu}}.
\ee
Here $\alpha$ is a label index running from 1 to 2, denoting fields corresponding to the two metrics/vielbeins, and $e_\alpha$ is shorthand for the determinant of the corresponding vielbein.
For ease of computation, in the following we use the Einstein equations obtained after a variation of the action with respect to the vielbeins and not the metrics, at the background cosmological level only, where the two versions are equivalent. They explicitly read
\be
G^{1\mu}_\nu = 8\pi G_N \beta_1 \frac{e}{e_1} (\frac{\beta_1}{2}(T^{\mu\lambda}g^{1}_{\lambda\nu}+g^{1}_{\nu\lambda}T^{\lambda\mu}) +\frac{\beta_2}{4} (T^{\mu\rho} Y_{\rho\nu}+Y_{\nu\rho} T^{\rho\mu})) +32 \pi G_N \Lambda^4 \frac{E^{1\mu}_a}{e_1} e_{1\nu}^a
\label{E1}
\ee
and
\be
G^{2\mu}_\nu = 8\pi G_N \beta_2 \frac{e}{e_2}( \frac{\beta_2}{2}(T^{\mu\lambda}g^2_{\lambda\nu}+g^{2}_{\nu\lambda}T^{\lambda\mu}) +\frac{\beta_1}{4}(T^{\mu\rho} Y_{\rho\nu}+Y_{\nu\rho} T^{\rho\mu}) ) +32 \pi G_N \Lambda^4 \frac{E^{2\mu}_a}{e_2} e_{2\nu}^a
\label{E2}
\ee
where we have used the symmetric vielbein condition explicitly.
We only make use of these equations in the background cosmological case where all the tensors are diagonal. In the general case, e.g. for cosmological perturbations,  these equations are not consistent as their antisymmetric parts are not guaranteed to vanish. In the background cosmological case, we will explicitly verify that the background solutions obtained with (\ref{E1}) and (\ref{E2}) coincide with the ones obtained  from the variation of the action with respect to the metrics. For the linear cosmological perturbations, we will use a more direct route and find the second order Lagrangian in each case explicitly and then derive the linear equations.
We have conveniently defined the duals
\be
E^{ia}_\mu = \epsilon^{\mu\nu\rho\sigma}\epsilon_{abcd} m^{ijkl} e^b_{j\nu} e^c_{k\rho} e^d_{l\sigma}
\ee
and the Jordan frame energy-momentum tensor
\be
T_{\mu\nu}= -\frac{2}{e} \frac{\delta S_m}{\delta g^{\mu\nu}}
\ee
which is obtained by varying the matter action with respect to the Jordan metric, i.e. not with respect to the two metrics $g^\alpha_{\mu\nu}$. This tensor plays a crucial role in the following.

\subsection{Cosmological background}

The previous Einstein equations at the background cosmological level can be specialised by choosing the cosmological ansatz for the metrics
\be
ds_1^2= a_1^2 (-N_1d\tau^2 +dx^2)
\ee
and
\be
ds_2^2= a_2^2 (- N_2 d\tau^2 +dx^2)
\ee
where the two lapse functions $N_{1,2}$ and the two scale factors $a_{1,2}$ differ.\footnote{This cosmological, FRW-like, ansatz is essentially the same as a mini-superspace ansatz.} We can always change to a unique conformal time by putting $d\eta= N_1 d\tau$ and introducing the ratio $b^2=\frac{N_2}{N_1}$ so that
\be
ds_1^2= a_1^2 (-d\eta^2 +dx^2)
\ee
and
\be
ds_2^2= a_2^2 (- b^2 d\eta^2 +dx^2)
\ee
where the ratio between the lapse functions $b^2$ plays a crucial role in the modification of gravity induced by the bigravity models.
We consider the coupling of bigravity to a perfect fluid defined by the energy-momentum tensor
\be
T^{\mu\nu}= (\rho+p) u^\mu u^\nu + p g^{\mu\nu}
\ee
where the 4-vector $u^\mu$ is
\be
u^\mu= \frac{dx^\mu}{d\tau_J}
\ee
and the proper time in the Jordan frame is simply
\be
d\tau_J^2 =- g_{\mu\nu}dx^\mu dx^\nu.
\ee
 We first consider the frame in which matter is at rest implying that
\be
T\equiv  g_{\mu\nu} T^{\mu\nu}= (-\rho+3p)
\ee
and $u^i=0$ at the cosmological background level, i.e. $g_{00} (u^0)^2=-1$ and therefore
\be
T^{00}=- g^{00} \rho.
\ee
Using the fact that
\be
ds^2= -(\beta_1 a_1 + \beta_2 b a_2)^2 d\eta^2 + (\beta_1 a_1 +\beta_2 a_2)^2 dx^2
\ee
we can identify the Jordan frame scale factor
\be
a_J= \beta_1 a_1 +\beta_2 a_2
\ee
and the conformal times
\be
d\eta_1 =d\eta,\ \ d\eta_2= b d\eta
\ee
when  the Jordan conformal time is
\be
d\eta_J= \frac{\beta_1 a_1 + \beta_2 b a_2}{\beta_1 a_1 +\beta_2 a_2}d\eta.
\ee
Matter is conserved in the Jordan frame, as follows from the residual diffeomorphism invariance (associated with diffeomorphisms of the Jordan frame metric) of the matter action,  implying that
\be
\frac{d\rho}{d\eta_J} + 3a_J { H}_J (\rho+p)=0
\ee
where the Jordan frame Hubble rate is identified with
\be
H_J\equiv \frac{d a_J}{a_J^2 d\eta_J}\equiv\frac{{\cal H}_J}{a_J}= \frac{1}{(\beta_1 a_1 + \beta_2 b a_2) a_J}(\beta_1 a_1^2 H_1  + \beta_2 a_2^2 H_2)
\ee
and we have introduced the two Hubble rates
\be
H_1=\frac{d a_1}{a_1^2 d\eta_1}\equiv \frac{d a_1}{a_1^2 d\eta},\ H_2=\frac{d a_2}{a_2^2 d\eta}.
\ee
When the  equation of state $\omega= \frac{p}{\rho}$ of the matter fluid is constant, we have that
\be
\rho= \frac{\rho_0}{a_J^{3(1+\omega)}}
\ee
where $\rho_0$ will be identified below.
We will also need the determinants
\be
e_1=a^4_1,\ \ e_2= ba_2^4, \ \ e= (\beta_1 a_1 + \beta_2 b a_2)(\beta_1 a_1 +\beta_2 a_2)^3
\ee
while we have the components of the vielbeins
\be
e^0_{10}= a_1, \  e^i_{1j}= a_1 \delta^i_j, \ e^0_{20}= a_2 b, \  e^i_{2j}= a_2 \delta^i_j.
\ee
The (00) component of Einstein's equations gives that
\be
G^{10}_{0}= -8\pi G_N  \beta_1 \frac{a_J^3}{a_1^3} \rho  -24\times 8\pi G_N \Lambda^4 \frac{a_1}{e_1} m^{1jkl}a_j a_k a_l.
\ee
where we have used $Y_{00}=-2\ b a_1 a_2$ and $E^{10}_0= -6 a_1 m^{1jkl}a_j a_k a_l$ as $\epsilon^{0abc}\epsilon_{0abc}=-6$.
Using $G^{10}_0= -3 H_1^2$, we get the Friedmann equation
\be
3H_1^2 m_{\rm Pl}^2=  \beta_1 \frac{a_J^3}{a_1^3}\rho +24\Lambda^4 m^{1jkl}\frac{a_j a_k a_l}{a_1^3}.
\label{F1}
\ee
Similarly we find that
\be
\frac{3H_2^2 m_{\rm Pl}^2}{b^2}=  \beta_2 \frac{a_J^3}{a_2^3}\rho +24\Lambda^4 m^{2jkl}\frac{a_j a_k a_l}{a_2^3}.
\label{F2}
\ee
We can also write the spatial components of the Einstein equations
\be
G^{1u}_{v}= 8\pi G_N \beta_1 \frac{e}{e_1} \frac{\beta_1 a_1^2 + \beta_2 a_1 a_2 }{a_J^2} p \delta^u_v + 8\pi G_N\times 24 \Lambda^4 m^{1jkl}\frac{\tilde a_j a_k a_l}{a_1^3} \delta^u_v
\ee
where we have used $Y_{uv}= 2a_1 a_2 \delta_{uv}$ and $E^{1u}_{v}= -6 m^{1jkl}a_1{\tilde a_j a_k a_l}\delta^u_v$. We have defined
\be
\tilde a_1= a_1, \ \tilde a_2= ba_2.
\ee
Now we have
\be
G^{1u}_{v}= (H_1^2 -2 \frac{1}{a_1^3} \frac{d^2 a_1}{d\eta_1^2}) \delta^u_v
\ee
implying the Raychaudhury equation
\be
2m_{\rm Pl}^2  \frac{1}{a_1^3} \frac{d^2 a_1}{d\eta^2}=m^2_{\rm Pl} H_1^2 - \beta_1 \frac{e}{e_1} \frac{\beta_1 a_1^2 +\beta_2  a_1a_2 }{a_J^2} p  + 24 \Lambda^4 m^{1jkl}\frac{\tilde a_j a_k a_l}{a_1^3}
\label{R1}
\ee
and similarly
\be
2m_{\rm Pl}^2  \frac{1}{a_2^3} \frac{d^2 a_2}{d\eta_2^2}=m^2_{\rm Pl} \frac{H_2^2}{b^2} - \beta_2 \frac{e}{e_2} \frac{\beta_2 a_2^2 + \beta_1 a_1 a_2 }{a_J^2} p +  24 \Lambda^4 m^{2jkl}\frac{\tilde a_j a_k a_l}{ba_2^3}.
\label{R2}
\ee
This implies the following differential equation for $b$
\be
2 \frac{H_2 m_{\rm Pl}^2}{a_2 }\frac{d\ln b}{d\eta}=2m_{\rm Pl}^2  \frac{1}{a_2^3} \frac{d^2 a_2}{d\eta^2} - H_2^2m_{\rm Pl}^2 + \beta_2 b^2 \frac{e}{e_2} \frac{\beta_2a_2^2 +\beta_1 a_1 a_2 }{a_J^2} p - 24 \Lambda^4 m^{2jkl}b . \frac{\tilde a_j a_k a_l}{a_2^3}.
\ee
This closes the system of equations describing the background cosmology of bigravity in FRW spaces when matter is a perfect fluid. Using the identity
\be
\frac{1}{a_2^3} \frac{d^2 a_2}{d\eta^2}= \frac{1}{a_2} \frac{dH_2}{d\eta} +2H_2^2
\ee
we finally find that
\begin{eqnarray}
 && \frac{H_2 m_{\rm Pl}^2}{a_2 }\frac{d\ln b}{d\eta}=\frac{m_{\rm Pl}^2}{a_2}\frac{dH_2}{d\eta}  +\frac{\beta_2 b^2}{2}\frac{e}{e_2}(  \frac{ba_2}{a_J} \rho  +  \frac{\beta_2 a_2^2 + \beta_1 a_1 a_2 }{a_J^2} p ) \nonumber \\ && + 12 \Lambda^4b  m^{2jkl} \frac{(ba_j-\tilde a_j) a_k a_l}{a_2^3}.\nonumber \\
\end{eqnarray}
We will analyse these equations below.

\subsection{The Bianchi identity}

Conservation of matter in the Jordan frame is ensured by the residual diffeomorphism invariance of the matter action (i.e. invariance of the matter action under diffeomorphisms acting on the Jordan metric, but not under separate diffeomorphisms for the two metrics) and implies that
\be
D_\mu T^{\mu\nu}=0
\ee
where $D_\mu$ is the covariant derivative associated to the Jordan frame metric. We will not use the explicit form of the conservation equation. On the other hand,  we will check directly that the two Friedmann equations \eqref{F1} and \eqref{F2} are compatible with the two Raychaudhuri equations \eqref{R1} and \eqref{R2}. This can be verified by directly taking the derivatives of the Friedmann equations with respect to $\eta_1$ and $\eta_2$ respectively. Using the first Friedmann and Raychaudhuri equations for instance, we find that they are compatible provided
\be
(1-\frac{a_2 H_2}{ba_1 H_1}) ( 24 \Lambda^4 m^{12kl}\frac{a_k a_l}{a_J^3} - \beta_1 \beta_2  p)=0.
\ee
This implies that the solutions exist on two different branches where either
\be
 24 \Lambda^4 m^{12kl}\frac{a_k a_l}{a_J^3}=\beta_1 \beta_2  p
 \label{b2}
\ee
or
\be
b=\frac{a_2 H_2}{a_1 H_1}.
\label{bi}
\ee
It can be explicitly checked that the second Raychaudhuri equation (\ref{R2}) is also compatible with the second Friedmann equation (\ref{F2}) when the conditions (\ref{bi},\ \ref{b2}) are satisfied. Hence we retrieve the fact that the background cosmology has two branches of solutions. In this paper, we will exclusively focus on the second branch \eqref{bi}.\footnote{There is an unfortunate clash of naming conventions for the two branches in the literature: The branch we consider in this paper is referred to as branch II in \cite{Comelli:2015pua,Lagos:2015sya}, but as branch I in \cite{Gumrukcuoglu:2015nua}.  The labels for branch I and II are therefore reversed between those sets of papers.}
When the condition (\ref{bi}) is applied, we find that the ratio between the scale factors $X= \frac{a_2}{a_1}$ is algebraically determined by the  time-dependent equation
\be
X= \frac{\beta_2 + \frac{24\Lambda^4}{\rho_0}(\beta_1 +\beta_2 X)^{3\omega} m^{2jkl} a_j a_k a_l}{\beta_1 + \frac{24\Lambda^4}{\rho_0}(\beta_1 +\beta_2 X)^{3\omega} m^{1jkl} a_j a_k a_l}
\ee
for which one can obtain two asymptotical regimes. When dark energy is negligible, i.e. in the radiation and matter eras, we have that
\be
X\to X_m= \frac{\beta_2}{\beta_1}
\ee
and in the asymptotic future when dark energy dominates we have that
\be
X\to X_d
\ee
 where
 \be
X_d= \frac{m^{2jkl} a_j a_k a_l}{m^{1jkl} a_j a_k a_l}.
\ee
We will come back to these eras when we describe the cosmological evolution of the model. In particular, we shall focus on the crucial role played by $b$ in these models.

\begin{figure*}
\centering
\includegraphics[width=0.49\linewidth]{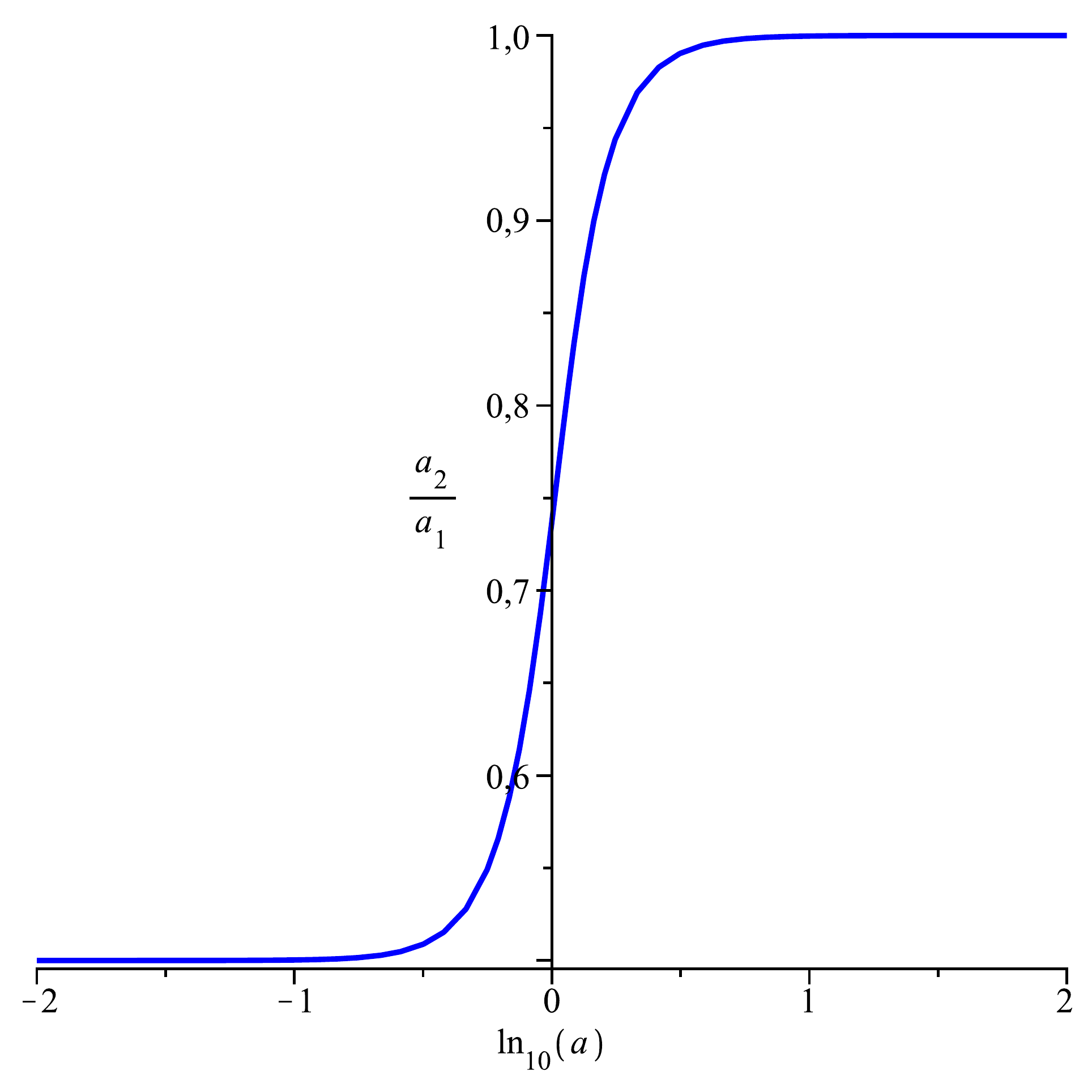}
\includegraphics[width=0.49\linewidth]{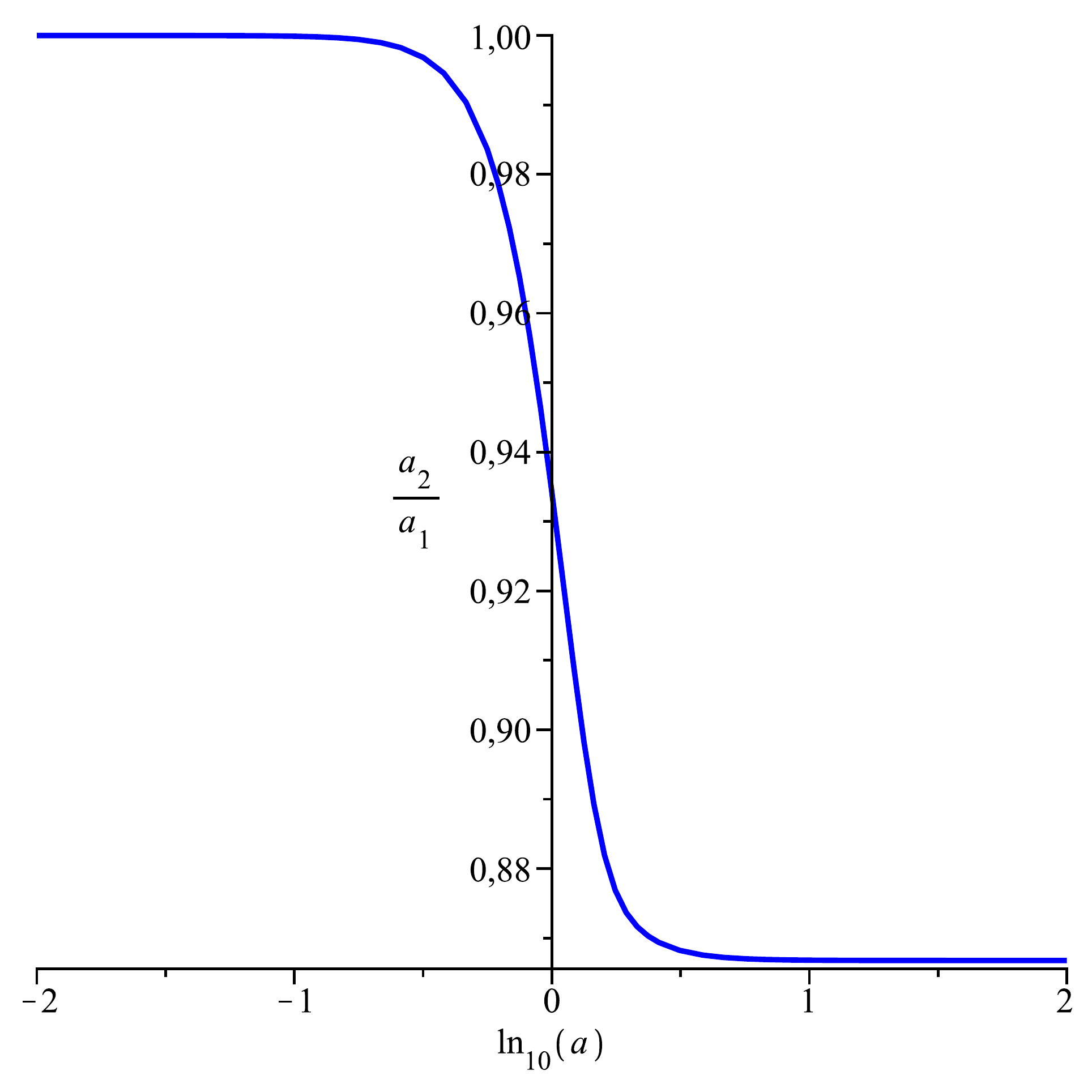}
\caption{The variation of $a_2/a_1$  as a function of  redshift $a_J$ from  $a_{\rm ini}=10^{-4}$ with  models I  (left panel) and model II (right panel) as described in sections \ref{subsecI} and \ref{subsecII}. One can easily see that $X$ goes from $X_m$ to $X_d$ between the matter era and the dark energy future.}
\end{figure*}

\begin{figure*}
\centering
\includegraphics[width=0.49\linewidth]{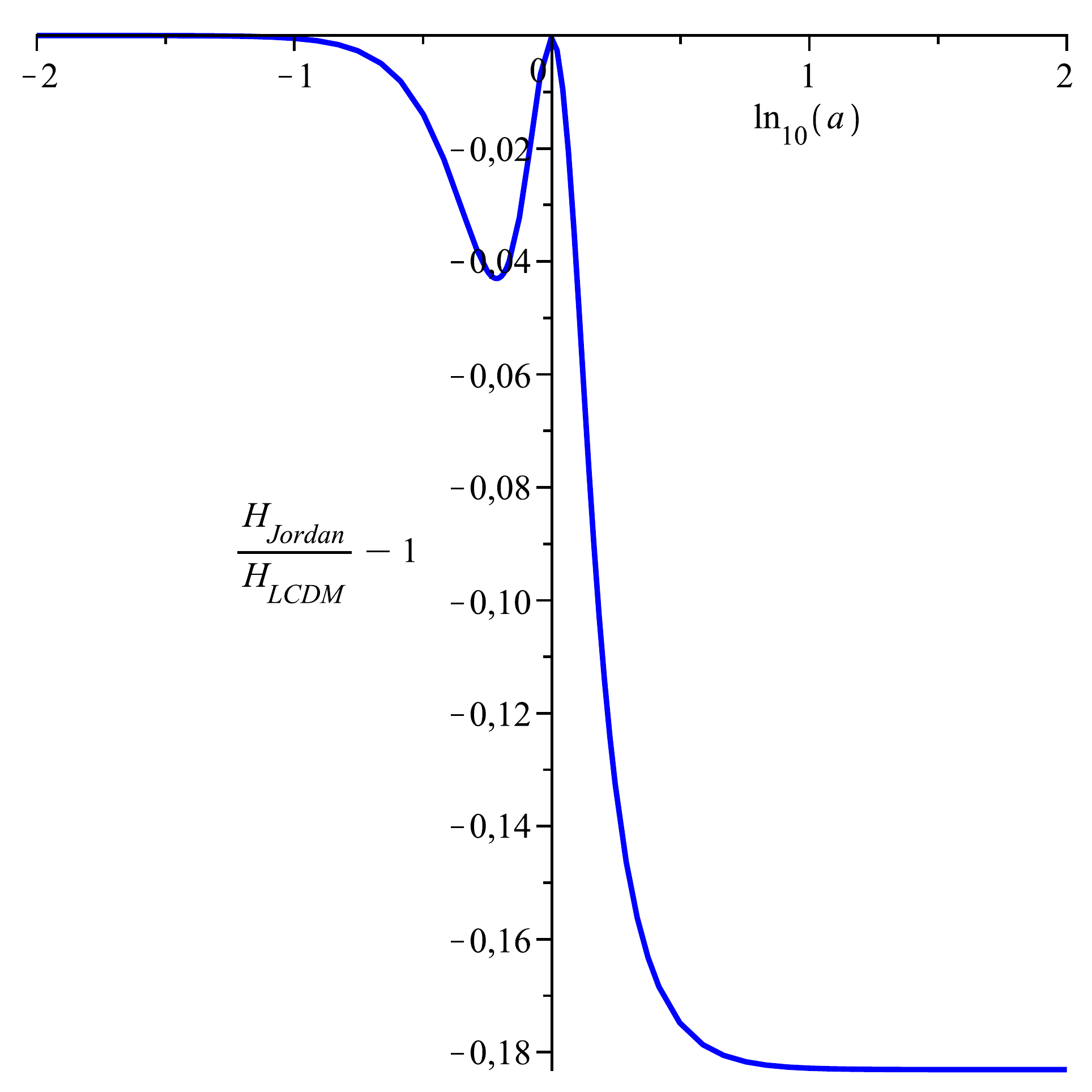}
\includegraphics[width=0.49\linewidth]{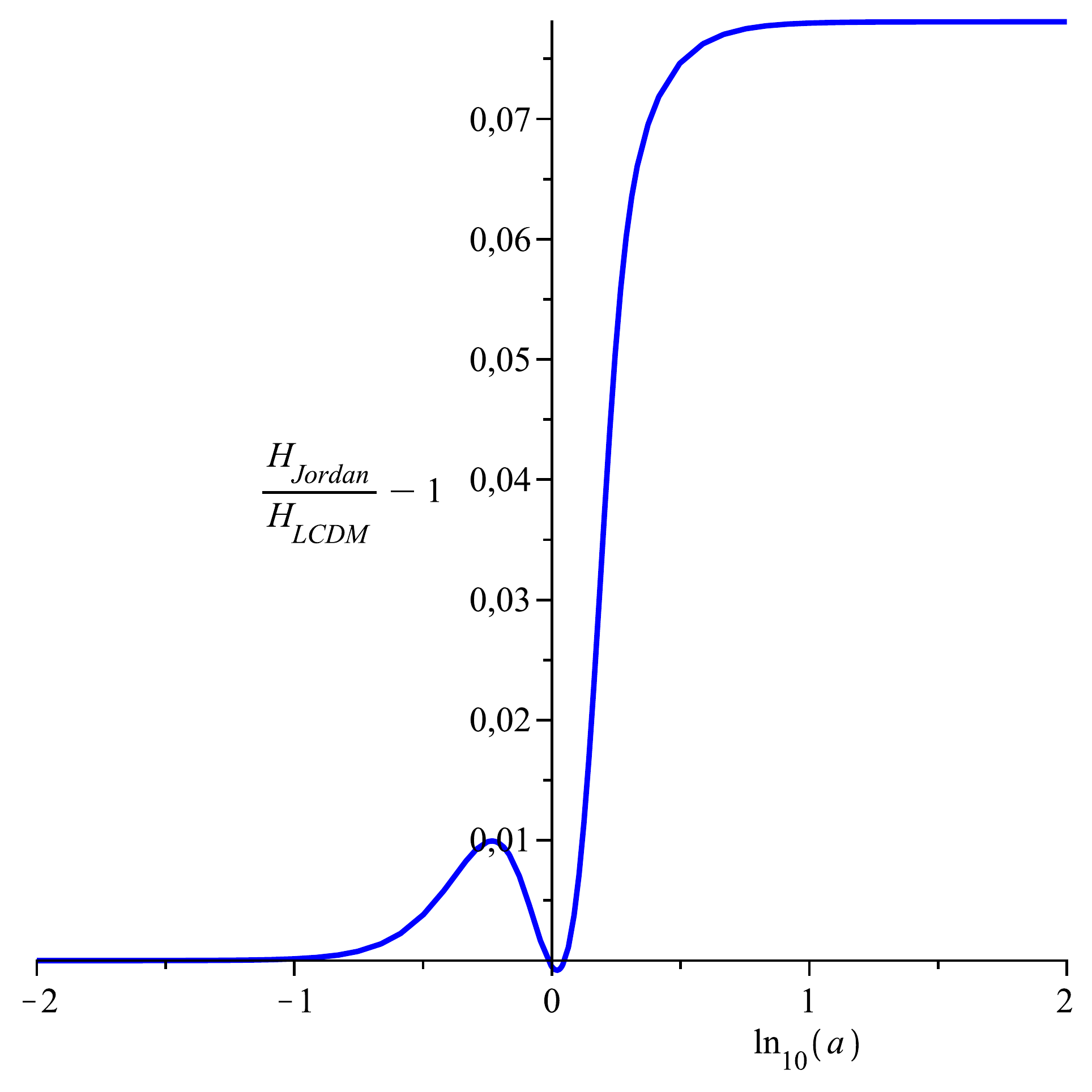}
\caption{The variation of $H_J/H_{\rm LCDM}-1$  as a function of  redshift $a_J$ from  $a_{\rm ini}=10^{-4}$ with  model I (left panel) and model II (right panel). The value of the coefficient $c$ (\ref{ci}) has been adjusted to have coincidence with $\Lambda$-CDM now. The asymptotic difference between the cosmological constant and the one of $\Lambda$-CDM is due to $c\ne 1$. }
\end{figure*}

\section{Scalar Cosmological perturbations}
\subsection{The GR case}

We are interested in linear cosmological perturbations around a flat cosmological background that we write in conformal coordinates. We will work with vielbeins as this is the formulation which will be extended to the bigravity case. Under a change of coordinates $x^\mu \to x^\mu + \xi^\mu$, the vielbeins transform as
\be
e^a_\mu \to e^a_\mu - \partial_\mu \xi^\nu e^a_\nu
\label{gauge}
\ee
and this can be used to reduce the number of degrees of freedom.
At the background level we have
\be
\bar e^a_\mu= a \delta^a_\mu
\ee
and we consider the most general scalar perturbations
\be
\delta e^0_0= a \Phi, \ \delta e^i_j = -a\Psi \delta^i_j + a\partial^i \partial_j U
\ee
 where the spatial index of the spatial derivative  is raised with $\delta^{ij}$, i.e. $\partial^i=\delta^{ij} \partial_j$ and
 \be
 \delta e^i_0= -a\partial^i W, \ \ \delta e^0_i= -a\partial_i V
 \ee
 comprising 5 degrees of freedom. Using the fact that $\delta g_{\mu\nu}=  \eta_{ab} (\bar e^a_\mu \delta e_\nu^b + \delta e^a_\mu \bar e_\nu^b)= a (\eta_{\mu b} \delta e^b_\nu + \eta_{a\nu} \delta e^a_\mu)$, we find explicitly that
 \be
 \delta g_{00}= - 2 a^2 \Phi, \ \ \delta g_{ij}= a^2(-2 \Psi \delta_{ij} + \partial_i\partial_j U)
 \ee
  and
  \be
  \delta g_{0i}=\delta g_{i0}= a^2 \partial_i (V-W).
  \ee
As a result only $(V-W)$ is a degree of freedom and we can choose $W=0$ to simplify the analysis. This implies that our ansatz now reads
\be
\delta e^0_0= a \Phi, \ \delta e^i_j = -a\Psi \delta^i_j + a\partial^i \partial_j U, \ \delta e^0_i= -a\partial_i V, \ \delta e^i_0=0
\ee
as a function of the four scalar degrees of freedom $(\Phi,\Psi, U, V)$.

 We can  use two gauge transformations with
\be
\xi^0= -V, \ \ \xi^i= \partial^i U
\ee
to gauge away the $U$ and $V$ scalars. Notice that this transformation would induce a variation of $e^i_0$ of the form $a\partial^i \partial_0 U$ and of $e^0_0$ like $a \partial_0 V$.  We use  the fact that we are only interested in the quasi-static regime where spatial derivatives dominate over time derivatives which are neglected in this regime. This condition is realised in the sub-horizon limit of cosmological perturbations where one studies perturbations on scales much smaller than the cosmological horizon, i.e. we only consider perturbations for which $k/a \gg H$. As a result  we find that the metric can be put in the conformal Newton gauge
\be
ds^2= a^2 (- (1+2\Phi) d\eta^2 + (1-2\Psi) dx^2)
\ee
and the Lagrangian comprising both the Einstein-Hilbert term and the coupling to pressure-less matter reads\footnote{In the flat spatial geometry, we use the vector notation $\partial_i = \vec \nabla$ and $(\vec\nabla a).(\vec\nabla b)= \delta^{ij}\partial_i a \partial_j b= \partial_i a \partial^i b$.}
\be
{\cal L}= \frac{a^2}{8\pi G_N} ((\vec\nabla \Psi)^2-2 \vec\nabla\Psi . \vec\nabla \Phi)
-a^4 \delta\rho \Phi
\ee
from which we deduce the unicity of the Newtonian potential
\be
\Phi=\Psi
\ee
 and the Poisson equation
\be
\Delta \Phi= 4\pi G_N a^2 \delta \rho.
\ee
We will generalise this analysis to the case of bigravity.

\subsection{Scalar perturbations in bigravity}

In the case of doubly-coupled bigravity, although most of our argument will go through unaltered in the singly-coupled case as well, we simply double the number of degrees of freedom prior to gauge fixing, i.e we have the two sets of scalars $(\Phi_{1,2},\Psi_{1,2}, V_{1,2}, U_{1,2})$.
{Recall that we also constrained our vielbeins (and accordingly also our perturbative ansatz in what follows) to satisfy the symmetric vielbein condition, which at the linear level implies}
\be
\eta_{ab}( \delta e^a_{1\mu} \bar e^{b}_{2\nu}- \delta e^b_{1\nu} \bar e^a_{2\mu})= \eta_{ab}( \delta e^a_{2\mu} \bar e^{b}_{1\nu}- \delta e^b_{2\nu} \bar e^a_{1\mu}).
\label{sym}
\ee
Notice that this is in fact dynamically implemented when considering the low energy/decoupling limit of the theory \cite{deRham:2015cha,Melville:2015dba}, although in general this has to be imposed separately if the equivalence with the metric formulation is to be guaranteed.
%
%
This also ensures the equivalence between the potential term in the metric and vielbein formalisms in our context. We then use this parameterisation of the perturbations as obtained in the appendix to couple them to matter at the Lagrangian level and eventually deduce their equations of motion. This symmetric vielbein condition imposes only one extra condition on the scalar perturbations which can be obtained using the $(0i)$ or $(i0)$ components and reads
\be
V_2= b V_1.
\label{sym}
\ee
This reduces the number of degrees of freedom to only seven.

In the bigravity case, only the diagonal subgroup of diffeomorphisms acting on both vielbeins is a symmetry of the theory. In the scalar sector, such gauge transformations are still specified by two scalar functions $\xi^\mu=(\xi^0, \partial^i \xi)$ which can remove only two scalar degrees of freedom, therefore reducing their number down to five.

More specifically, we can use two gauge transformations with respectively
\be
\xi^0=-{V_1}= -\frac{V_2}{b}
\ee
and
\be
\xi^i= \partial^i U_2
\ee
to gauge away $(V_1, V_2, U_2)$. Indeed we can check that we have explicitly
\be
\delta e_{1i}^0= -a_1 \partial_i V_1\ \to \ \delta e_{1i}^0-\partial_i \xi^0 \bar e_{10}^0=-a_1 \partial_i V_1+a_1 \partial_i V_1=0
\ee
and similarly
\be
\delta e_{2i}^0= -a_2 \partial_i V_2\ \to \ \delta e_{2i}^0-\partial_i \xi^0 \bar e_{20}^0=-a_2 \partial_i V_2+a_2 b \partial_i V_1=0
\ee
where we have used (\ref{sym}) explicitly. The cancellation of $\delta e^i_{2i}$ works in a similar manner indeed
we have
\be
\delta e_{2i}^j= -a_2 \Psi_2 +a_2 \partial^j\partial_i U_2\ \to\ \delta e_{2i}^j- \partial_i\xi^k\bar e_{2_k}^j= -a_2 \Psi_2
\ee
 and finally
\be
\delta e_{1i}^j= -a_1 \Psi_1 +a_1 \partial^j\partial_i U_1\ \to\ \delta e_{1i}^j- \partial_i\xi^k\bar e_{1_k}^j= -a_1 \Psi_1+a_1 \partial^j\partial_i (U_1-U_2).
\ee
After these gauge transformations
we are thus left with five degrees of freedom in the gravitational sector $(\Phi_{1,2}, \Psi_{1,2}, U)$ where we have defined
\be
U=U_1-U_2
\ee
and the perturbations are defined by
\be
\delta e^0_{10}= a_1 \Phi_1, \ \delta e^i_{1j} = -a_1\Psi_1 \delta^i_j + a_1\partial^i \partial_j U, \ \delta e^0_{1i}=0, \ \delta e^i_{10}=0
\ee
and
\be
\delta e^0_{20}= a_2 \Phi_2, \ \delta e^i_{2j} = -a_2\Psi_2 \delta^i_j, \ \delta e^0_{2i}=0, \ \delta e^i_{20}=0.
\ee
The previous results are valid in the quasi-static approximation which can be implemented in the perturbative regime on sub-horizon scales such that $H\ll k/a \lesssim \Lambda_3$.
We will analyse the dynamics of bigravity when these perturbations are turned on.

\subsection{The Poisson equations}
\label{sec:poisson}

We have to write down the Einstein-Hilbert terms and the potential when the perturbations are present. We focus only on the quasi-static regime in order to generalise the GR derivation of the Poisson equation.
The Einstein Hilbert term for the second metric $g^2_{\mu\nu}$ coincides with the one of GR in the conformal Newtonian gauge. Let us now examine the one of the first metric $g^1_{\mu\nu}$. For that we
will use the fact that Einstein-Hilbert term is invariant under reparametrisation and therefore one can formally gauge away $U$. Hence the Einstein-Hilbert term of the first metric is independent of $U$. It will prove useful to absorb the trace part of $\partial^i \partial_j U$ in the Newtonian potential $\Psi_1$ by redefining
\be
\delta e^i_{1j} = a(-\tilde \Psi_1 \delta^i_j + \partial^i \partial_j U- \frac{\Delta U}{3} \delta^i_j).
\label{an}
\ee
With this field redefinition the Einstein-Hilbert terms of the model lead to the Lagrangian
\be
{\cal L}_{EH}(\tilde\Psi_1,\Psi_2,\Phi_{1,2},U)= \frac{a_1^2}{8\pi G_N} (((\vec\nabla (\tilde \Psi_1- \frac{\Delta U}{3} ))^2-2 \vec\nabla(\tilde \Psi_1 -\frac{\Delta U}{3}) .\vec\nabla \Phi_1)+ \frac{b a_2^2}{8\pi G_N} ((\vec\nabla \Psi_2)^2-2 \vec\nabla\Psi_2  \vec\nabla \Phi_2).
\ee
which is also
\be
{\cal L}_{EH}(\Psi_{1,2},\Phi_{1,2})= \frac{a_1^2}{8\pi G_N} ((\vec\nabla \Psi_1 )^2-2 \vec\nabla\Psi_1 .\vec\nabla \Phi_1)+ \frac{b a_2^2}{8\pi G_N} ((\vec\nabla \Psi_2)^2-2 \vec\nabla\Psi_2  \vec\nabla \Phi_2).
\label{an1}
\ee
when reverting to $\Psi_1$. But let us work with the parametrisation (\ref{an}) first. In this case the
 new terms coming from the potential at second order are either algebraic in $(\Phi_{1,2},\tilde\Psi_1,\Psi_2)$ or involve one or two terms in $(\partial^i \partial_j U- \frac{\Delta U}{3} \delta^i_j)$.
 The algebraic terms at second order are mass terms of order $ \frac{\Lambda^4}{m_{\rm Pl}^2}=m^2\sim H_0^2$ for the four potentials. As we work in the subhorizon limit where spatial derivatives are much larger than the Hubble rate, these terms are negligible compared to the Einstein-Hilbert terms which act as kinetic terms for the four potentials $(\Phi_{1,2},\tilde\Psi_1,\Psi_2)$.
The mass term for the Newton potentials would lead to a Yukawa suppression of the potentials on large scales of the form $e^{-m r}$ which is negligible for distances $r \ll H_0^{-1}$ where we apply the Newtonian analysis followed here. For a more complete discussion in the local Minkowski limit, see section \ref{sec:local}.  On the other hand on horizon scales, we would not be able to use  this approximation anymore.

  As a result we will neglect the algebraic terms coming from the potential of bigravity.
The terms involving $U$ give rise to new kinetic terms  and we shall focus on them here.
Let us first deal with terms linear in $(\partial^i \partial_j U- \frac{\Delta U}{3} \delta^i_j)$. As the other components of the vielbeins are all diagonal elements we get terms like
\be
\epsilon_{0abc} \epsilon^{0 ijk} (\partial^a \partial_i U- \frac{\Delta U}{3} \delta^a_i)\delta^b_j\delta^c_k\propto \delta^i_a(\partial^a \partial_i U- \frac{\Delta U}{3} \delta^a_i)=0
\ee
hence all the terms linear in $U$ cancel. We are left with terms involving two powers of $U$. They look like
\be \label{Uterm}
\epsilon_{0abc} \epsilon^{0 ijk} (\partial^a \partial_i U- \frac{\Delta U}{3} \delta^a_i)(\partial^b \partial_j U- \frac{\Delta U}{3} \delta^b_j)\delta^c_k\propto
(\partial^j \partial_i U- \frac{\Delta U}{3} \delta^j_i)(\partial^i \partial_j U- \frac{\Delta U}{3} \delta^i_j).
\ee
These terms are higher order kinetic terms for the field $U$, which is completely decoupled at second order in perturbations in the quasi-static approximation from both the four Newtonian
potentials $(\Psi_{1,2},\Phi_{1,2})$ and matter. Indeed the structure of the Lagrangian in the quasi-static and sub-horizon limit comprises three terms
 \be
 {\cal L}= {\cal L}_{EH}(\tilde\Psi_1,\Psi_2,\Phi_{1,2},U) + {\cal L}_U + {\cal L}_m
\ee
where ${\cal L}_U \propto \Lambda^4 m_{11} (\partial^j \partial_i U- \frac{\Delta U}{3} \delta^j_i)(\partial^i \partial_j U- \frac{\Delta U}{3} \delta^i_j)$ and the matter Lagrangian, when only pressure-less matter is involved,
couples the two potential $\Phi_{1,2}$ to the matter density (see below). We can now perform a field redefinition going back to $\Psi_1= \tilde \Psi_1 -\frac{\Delta U}{3}$ and write
 \be
 {\cal L}= {\cal L}_{EH}(\Psi_{1,2},\Psi_2,\Phi_{1,2}) + {\cal L}_U + {\cal L}_m
\ee
which proves that in the quasi-static and sub-horizon limit when matter is pressure-less, the $U$ field decouples from the dynamics of perturbations completely and can be discarded.
This comes from the fact that pressure-less matter only couples to $\Phi_{1,2}$ and not $\Psi_{1,2}$.
Nonetheless, the $U$ field has an action of higher order in its  derivatives of the form $U\Delta^2 U$ which is not of the Galileon type nor a total derivative and is therefore the signal that, if we went beyond the quasi-static approximation, {thus restoring the corresponding higher-order time-derivatives}, the $U$ field would {give rise to} a ghost in the theory.

Explicitly demonstrating that $U$ would give rise to a ghost-like degree of freedom and in fact propagates two scalar degrees of freedom is straightforward. Going back beyond the quasi-static approximation we restore time-derivatives in the Minkowski limit and promote \eqref{Uterm} to
\begin{align}
{\cal L_U} \propto  \Lambda^4 \int d^4x \Box U \Box U . \label{Uterm2}
\end{align}
 where we have integrated by parts and covariantised $\Delta$ to a full 4D D'Alembertian $\Box$. We can now rewrite this interaction in the following way
\be
 \Lambda^4 \int d^4x \Box U \Box U \to   \int d^4x \left( \Lambda^3 X \Box U - \frac{\Lambda^2}{4}X^2 \right).
\ee
The $U$ field has mass dimension $[U]=-2$.
We have introduced the auxiliary field $X$ in the first line whose dimension is one $[X]=1$. The action for $U$ and $X$ is dynamically equivalent  to (\ref{Uterm2}) after substituting the equation of motion  $X=2 \Lambda^2 \Box U $. It is convenient to redefine $\bar U=\Lambda^3 U$ whose dimension is $[\bar U]=1$. The resulting action is then
\be
\int d^4x \left(  X \Box \bar U - \frac{\Lambda^2}{4}X^2 \right)
\ee
We then diagonalise the kinetic terms by replacing $\bar U \to {\hat U + \hat X}$ and $X \to {\hat X - \hat U}{}$. The resulting action
\be
\int d^4x \left( \hat X \Box \hat X - \hat U \Box \hat U -\frac{\Lambda^2}{4}\hat X^2 -\frac{\Lambda^2}{4}\hat U^2 + \frac{\Lambda^2}{2}\hat X \hat U  \right).
\ee
clearly describes two dynamical second-order scalar degrees of freedom with opposite sign kinetic terms with a mixing mass matrix. This demonstrates that one recovers one ghost and one healthy scalar from the original U interactions. For additional details see the related discussion in section 8 of \cite{Noller:2013yja}.

We can also consider the coupling to matter of both the transverse traceless graviton in the Jordan frame and the $U$ field which reads
\be
\int d^4x \left (\frac{ \bar h_{ij}}{m_{\rm Pl}} + \frac{\beta_1}{\Lambda^3} \partial_i \partial_j \bar U\right )T^{ij}
\label{coup}
\ee
where $T^{ij}$ is the spatial part of the energy momentum tensor in the Jordan frame and $\bar h$ has dimension one. After the change of field and the introduction of the normalised pair $(\hat X,\hat U)$ this becomes
\be
\int d^4x \left (\frac{\bar h_{ij}}{m_{\rm Pl}} + \frac{\beta_1}{\Lambda^3} ( \partial_i \partial_j \hat U +\partial_i\partial \hat X)\right )T^{ij}.
\ee
Notice that this is the coupling that one expects with a two derivative interaction suppressed by the scale $\Lambda$.

The mass matrix of $(\hat X,\hat U)$ has a zero eigenvalue corresponding to the  massless excitation $\bar U$ while $X$ has a mass  $\Lambda$.  At low energy below $\Lambda$,  the field $X$ can be integrated out and we retrieve a massless scalar field $\bar U$ with a higher order kinetic term
\begin{align}
{\cal L_{\bar U}} \propto   \int d^4x \frac{\Box \bar U \Box\bar  U}{\Lambda^2} . \label{Uterm3}
\end{align}
and a derivative coupling (\ref{coup}) to matter.

The above is similar to the result of \cite{Gumrukcuoglu:2015nua} where the same degree of freedom was shown to give rise to a ghost in the late time Universe. Its presence requires further investigation but here at the linear level of cosmological perturbations {and in the quasi-static approximation}, we simply acknowledge that U decouples from matter. Note, however, that one may expect this scalar ghost to be a remnant of the ghost-like degree of freedom that propagates in doubly-coupled models at energy scales beyond the $\Lambda_3^3=m_{\rm Pl} m^2$ decoupling limit \cite{deRham:2014naa} { and hence to be harmless}. This is suggested by the previous analysis in terms of the fields $(\hat X, \hat U)$  where the ghost field acquires a mass of order $\Lambda \gg \Lambda_3$. A proper analysis of whether this is in fact the case would involve integrating out the ghost and other interaction terms above the scale $\Lambda_3$ in order to systematically investigate the resulting low-energy theory.  Again we will leave this for further investigation.

Let us summarise our result and ask ourselves when the decoupling of $U$ is guaranteed. This decoupling operates in the sub-horizon limit which allowed us to neglect the mass terms for the $(\Phi_{1,2},\tilde\Psi_1,\Psi_2)$ fields. One can expect that a more general treatment involving all the perturbations should be necessary on large horizon scales. We have also used the quasi-static approximation to gauge away some of the degrees of freedom such as $U_2$ and this assumption should also be revised in situations where time derivatives could compete with spatial gradients. Moreover we have assumed that linear perturbation theory is valid. This is certainly valid cosmologically for the Newtonian potentials which can only reach values of order $10^{-4}$ for large galaxy clusters. We can also use the present approach in the static situation corresponding to the solar system. In these cases, the quasi-static and sub-horizon approximation apply whilst the Newtonian potentials do not exceed the one of the sun, i.e. around $10^{-6}$. As a result, we will safely neglect the $U$ field in local gravitational cases. This will allow us to calibrate Newton's constant to the local one (see below). On the other hand, our approach would certainly fail in the strong gravitational regime of neutron stars or black holes.

\subsection{Scalar perturbative dynamics}
The cosmological perturbations involve tensor, vector and scalar modes.  In this section, we will exclusively concentrate on the scalar modes as they have a direct influence on the growth of structure. We have seen that the cosmological dynamics in the quasi-static limit reduces to the evolution of four Newtonian potentials $(\Phi_{1,2},\Psi_{1,2})$. In the Jordan frame where matter couples minimally to the Jordan metric, the matter perturbations are described by the fluid velocity $\vec{v}$ and the matter density contrast $\delta = \frac{\delta \rho}{\rho}$. The metric perturbations in the Jordan frame reduce to two Newtonian potentials $\Phi_J$ and $\Psi_J$ which govern the behaviour of matter and photon geodesics. In the following,
we will only be interested in the sub-horizon limit of perturbations  where $k/a_J \ll H_J$ and situations where the linear approximation for the gravitational potentials is valid $\vert \Psi_J\vert \ll 1, \ \ \vert\Phi_J\vert \ll 1$. In the Jordan frame, the matter particles behave like a fluid with velocity
$\vec{v}$ which follows the geodesics of the Jordan metric $g_{\mu\nu}$. The equations of motions for this fluid follow uniquely from conservation of matter in the Jordan frame, i.e. there is no need to incorporate the Einstein equation to find the equations of motion for the fluid.

In order to find the relationship between the Newtonian potentials $\Psi_J$ and $\Phi_J$ in the Jordan frame and matter, i.e. the new Poisson equations, we {use} the four Newtonian potentials $(\Psi_{1,2},\Phi_{1,2})$, where the fifth degree of freedom $U$ decouples in the sub-horizon and quasi-static approximation as discussed in the previous section. Two of the remaining degrees of freedom will turn out to be spurious, i.e. we will end with only two dynamical Poisson equations. Eventually when one takes into account the matter perturbation, i.e. the density contrast, in the scalar sector we end up with three scalar perturbations.  For this,
let us first define the perturbed metrics
\be
ds_1^2= a_1^2 (-(1+2\Phi_1)d\eta^2 + (1-2\Psi_1) dx^2)
\ee
and
\be
ds_2^2= a_2^2 (- b^2(1+2\Phi_2)d\eta^2 + (1-2\Psi_2) dx^2)
\ee
from which we can read off the {constrained} vielbeins
\be
e^0_{10}=a_1(1+\Phi_1), \ \ e^{u}_{1v}= a_1 (1-\Psi_1) \delta^u_v
\ee
and
\be
e^0_{20}=a_2b(1+\Phi_2), \ \ e^{u}_{2v}= a_2 (1-\Psi_2) \delta^u_v.
\ee
The Jordan frame vielbeins become
\be
e^0_0= (1+\Phi_J) \bar e^0_0,\ e^u_v= (1-\Psi_J) \bar e^u_v
\ee
where
\be
\bar e^0_0= \beta_1 a_1 +\beta_2 a_2 b,\ \ \bar e^u_v= a_J \delta^u_v
\ee
and we find the two potentials in the Jordan frame
\be
\Phi_J= \frac{\beta_1 a_1 \Phi_1 + \beta_2 a_2 b \Phi_2}{\beta_1 a_1 +\beta_2 a_2 b}, \ \ \Psi_J= \frac{\beta_1 a_1 \Psi_1 + \beta_2 a_2  \Psi_2}{\beta_1 a_1 +\beta_2 a_2 b}.
\ee
Geodesics are influenced by the gravitational force $-\nabla \Phi_J$ while light rays respond to $(\Phi_J + \Psi_J)/2$.
In the presence of a matter overdensity $\delta \rho$, the Poisson equations read
\be
\Delta \Phi_J = 4\pi G_N^\Phi a_J^2 \delta \rho , \ \ \Delta \Psi_J = 4\pi G_N^\Psi a_J^2 \delta \rho.
\ee
It is conventional to introduce different combinations of these Newton constants.
First of all, one can define the slip parameter $\eta$ which measures how much the two potentials differ. When the two potentials differ, this is a clear modification of gravity as we have seen that in GR the two potentials are equal. The slip parameter $\eta$ is defined by
\be
\eta\equiv \frac{\Psi_J}{\Phi_J} = \frac{G_N^\Psi}{G_N^\Phi}
\ee
and it differs from one generically (see below). When the gravitational acceleration $-\vec\nabla \Phi_J$ differs from the Newtonian acceleration $-\vec\nabla \Phi_N$ where $\Phi_N$ is the Newtonian potential defined in the section on local dynamics (section 5), structures grow at a different rate because of the modified gravitational interaction. This can be captured by
defining
\be
\mu\equiv \frac{G_N^\Phi}{G_N^{\rm local}}
\ee
where $G_N^{\rm local}$ is the local Newton constant in Minkowski space which will be identified below.
When this is not equal to one, this  implies a modification of the growth of structure. Finally we introduce a parameter $\Sigma$ which quantifies how much lensing by dark matter is going to be affected by
a modification of gravity
\be
\Sigma\equiv \frac{G_N^\Phi+G_N^\Psi}{2G_N^{\rm local}}= \mu \frac{(1+\eta)}{2}
\ee
which will not be equal to one either and therefore lensing will be affected. Indeed, this follows from the link between the  lensing potential and matter
\be
\Phi_W= \frac{\Phi_J+ \Psi_J}{2}
\ee
given by the Poisson equation
\be
\Delta \Phi_W= 4\pi G_N^{\rm local} a^2_J  \Sigma\delta \rho.
\ee
The Poisson equation which influences the growth of structure reads
\be
\Delta \Phi_J= 4\pi G_N^{\rm local} a^2_J  \mu \delta \rho
\ee
where $\mu$ will be determined below.

The conservation of matter and the Euler equation are  not modified in the Jordan frame, this follows from the residual diffeomorphism invariance of the matter action. They read
\be
\frac{\partial\delta}{\partial \eta_J}+ \partial_i v^i=0
\ee
and
\be
\frac{\partial v^i}{\partial \eta_J} + {\cal H}_J v^i= -\partial^i \Phi_J
\ee
where we have used $u^\mu=a_J^{-1} (1-\Phi_J+ v_i v^i, v^i)$. Here $v^i$ is the velocity of the matter particles and indices are lowered with $\delta_{ij}$. The gradient $\partial_i= \frac{\partial}{\partial x^i}$ is the comoving one. This allows one to
 deduce the growth equation for the
density contrast
\be
\frac{\partial^2 \delta }{\partial\eta_J^2} + {\cal H}_J \frac{\partial\delta}{\partial\eta_J} -\frac{3}{2} \Omega_m \mu {\cal H}^2_J \delta =0.
\ee
where we have defined ${\cal H}_J= \frac{d a_J}{a_J d\eta_J}$ and $8\pi G_{N}^{\rm local}a_J^2 \rho= 3 \Omega_m {\cal H}_J^2$ is the matter fraction.
As soon as $G_N^\Phi$ and/or the background cosmology is not the one of $\Lambda$-CDM, the growth of structure is modified.

\subsection{The Newtonian potentials}

It is transparent to deduce the equations of motion of the Newtonian potentials directly from the action of the model using the particular ansatz for the metrics and vielbeins that we have already discussed, see also the appendix.
The quadratic expansion of the Lagrangian involves mass terms for the potentials $\Phi_{1,2}$ and $\Psi_{1,2}$ of order $\Lambda^4/ m_{\rm Pl}^2 \sim m^2 \sim H_0^2$. We consider perturbations in the sub horizon limit where
$k/a_J \gg H_0$, implying that one can neglect the influence of these mass terms on the Newtonian potentials.
We can expand the Lagrangian to obtain
\bea
&&{\cal L}= \frac{a_1^2}{8\pi G_N} ((\vec\nabla \Psi_1)^2-2 \vec\nabla\Psi_1 . \vec\nabla \Phi_1)+ \frac{b a_2^2}{8\pi G_N} ((\vec\nabla \Psi_2)^2-2 \vec\nabla\Psi_2 . \vec\nabla \Phi_2)
\nonumber \\ && +e \bar g^{00} \delta\rho (\beta_1 a_1 \Phi_1+ \beta_2 a_2 b \Phi_2) (\beta_1 a_1 + \beta_2 b a_2)\nonumber \\
\eea
where we consider only pressure-less fluids like Cold Dark Matter (CDM) and $\bar g^{00}=-(\beta_1 a_1 + \beta_2 b a_2)^{-2}$.
The Euler-Lagrange equations for $\Psi_{1,2}$ read
\be
\Delta (\Psi_{1,2}-\Phi_{1,2})=0.
\ee
As a result we find that each of the metrics depends on only one potential
\be
\Psi_{1,2}=\Phi_{1,2}
\ee
and we have the two Poisson equations
\be
\Delta\Phi_1= - 4\pi G_N \frac{e}{e_1} (a_1^2 \bar g^{00}) \beta_1 a_1 (\beta_1 a_1 + \beta_2 b a_2)\delta \rho
\ee
and
\be
\Delta\Phi_2= - 4\pi G_N \frac{e}{e_2} (a_2^2 \bar g^{00}) \beta_2 a_2 (\beta_1 a_1 + \beta_2 b a_2)\delta \rho
\ee
from which we can read off the growth parameter
\be
\mu\equiv \frac{G_N^\Phi}{G_N^{\rm local}}= -\frac{\frac{e}{e_1}(a_1^2 \bar g^{00}) \beta_1^2 a_1^2 + \frac{e}{e_2}(a_2^2 \bar g^{00}) \beta_2^2 a_2^2 b}{(\beta_1 a_1  +\beta_2 a_2)^2}\frac{G_N}{G_N^{\rm local}}
\ee
and
\be
\frac{G_N^\Psi}{G_N}= -\left( \frac{\frac{e}{e_1}(a_1^2 \bar g^{00}) \beta_1^2 a_1^2 + \frac{e}{e_2}(a_2^2 \bar g^{00}) \beta_2^2 a_2^2 }{(\beta_1 a_1  +\beta_2 a_2)^2}\right )
\left( \frac{\beta_1 a_1 + \beta_2 a_2 b}{\beta_1 a_1 + \beta_2 a_2 }\right ).
\ee
The two potentials only differ when $b\ne 1$. In particular we have for the slip function
\be
\eta=
\left( \frac{\frac{e}{e_1}(a_1^2 \bar g^{00}) \beta_1^2 a_1^2 + \frac{e}{e_2}(a_2^2 \bar g^{00}) \beta_2^2 a_2^2 }{\frac{e}{e_1}(a_1^2 \bar g^{00}) \beta_1^2 a_1^2 + \frac{e}{e_2}(a_2^2 \bar g^{00}) \beta_2^2 a_2^2 b }\right) \left(\frac{\beta_1 a_1 + \beta_2 a_2 b}{\beta_1 a_1 + \beta_2 a_2 }\right).
\ee
Notice that the slip $\eta$ is always equal to one when $b=1$. We will see that this occurs in the matter-radiation and dark energy eras.
\begin{figure*}
\centering
\includegraphics[width=0.49\linewidth]{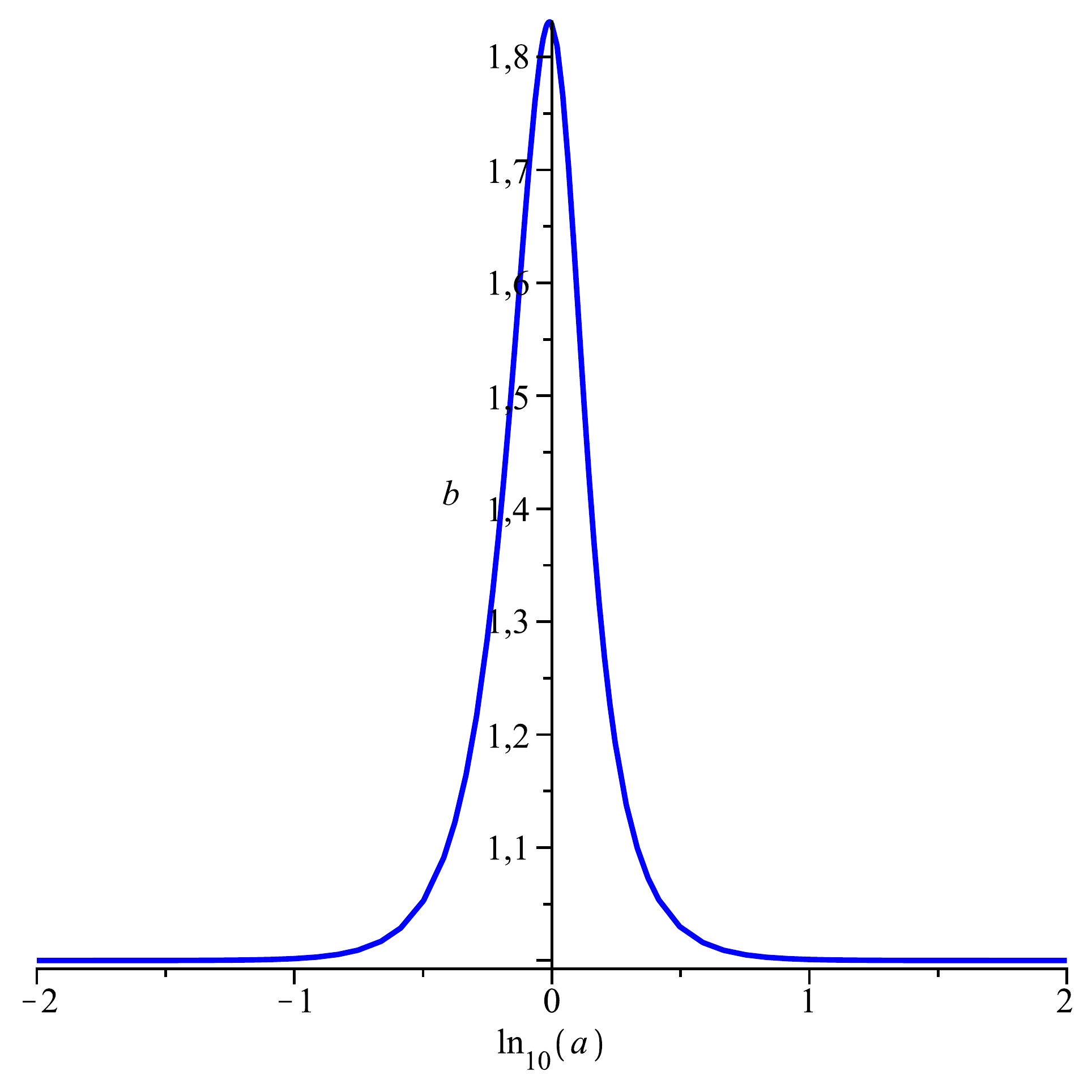}
\includegraphics[width=0.49\linewidth]{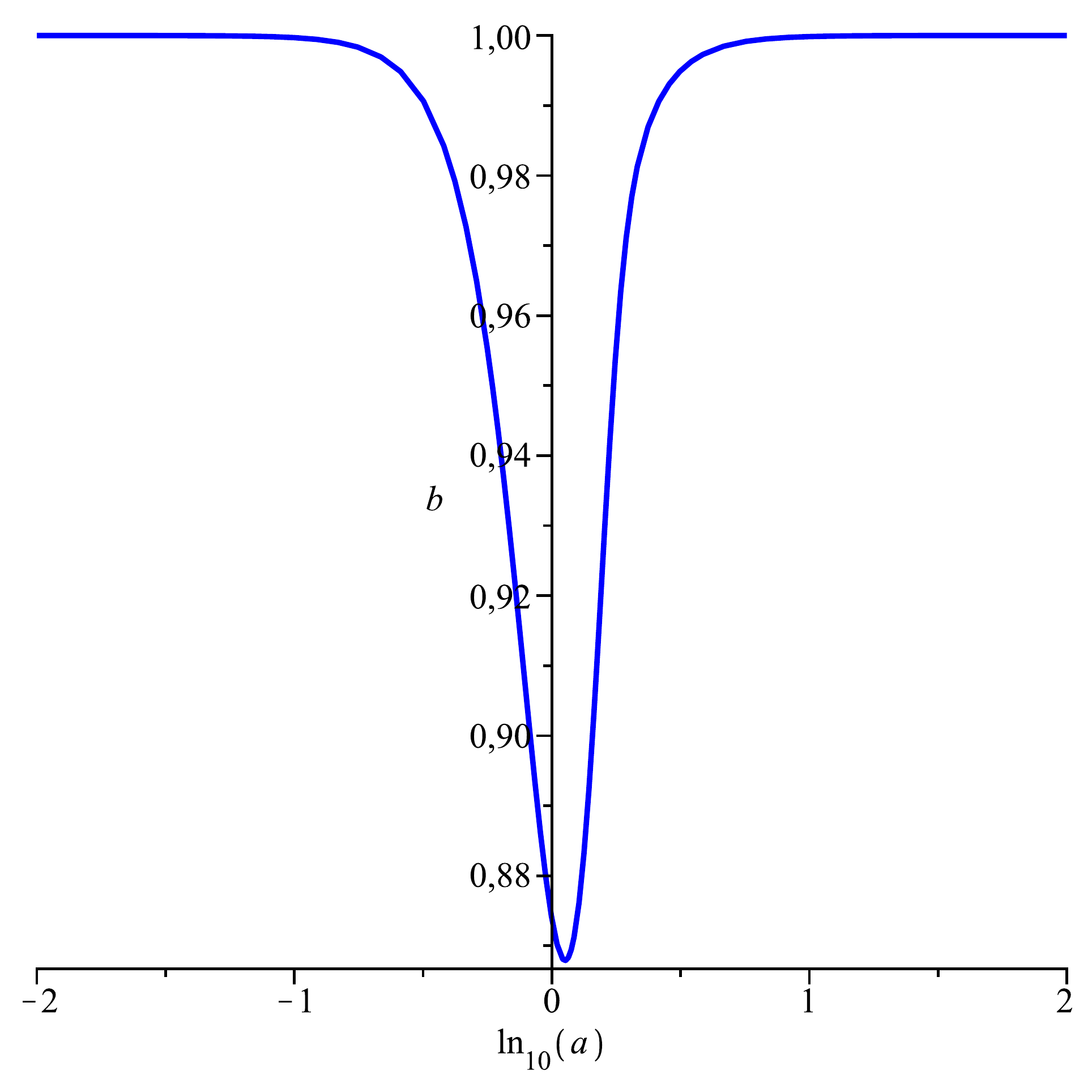}
\caption{The variation of the lapse function $b$  as a function of  redshift $a_J$ from  $a_{\rm ini}=10^{-4}$ with model I (left panel) and model II (right panel). The variation is only present between the two asymptotic regions. }
\end{figure*}

{

\subsection{Vector and tensor perturbations}

\subsubsection{Vector perturbations}

{The description of the vector degrees of freedom is given explicitly in the appendix.} We repeat the main results here for convenience. The most general vector perturbations in the vielbein formalism read
{
\be
\delta e^{i\alpha }_j =  a_\alpha (\partial_j V^i_\alpha +\partial^i W_{j\alpha})
\ee
 where the spatial index of the spatial derivative  is raised with $\delta^{ij}$, i.e. $\partial^i=\delta^{ij} \partial_j$, the index $\alpha=1,2$  and
\be
\delta e^{i\alpha}_0=  a_\alpha D^i_\alpha, \ \ \delta e^{0\alpha}_i= a_\alpha C_{i\alpha}.
\ee
The transversality conditions on these vectors in the scalar-vector-tensor decomposition are
\be
\partial^i C_{i\alpha}=0, \ \partial_i D^i_\alpha=0,\ \partial_i V^i_{\alpha}=0,\ \partial^i W_{i\alpha}=0.
\ee
The fact that $\delta g^\alpha _{0i}$ only involves $-b_\alpha C_i^\alpha + D^\alpha_i$, where $b_1=1$ and $b_2=b$,  allows us to  {choose the gauge such that} $D^i_\alpha=0$. {Indeed we use the vielbein formalism subject to the symmetric condition (\ref{sym}) and therefore the action depends on the two metrics $g^\alpha _{\mu\nu}$ only. } Similarly, as $\delta g^\alpha_{ij}$ depends only on $V_{i\alpha}+W_{i\alpha}$,
this allows us to choose $W_{i\alpha}=V_{i\alpha}$.
Now the symmetric condition  on the vielbeins implies also that
\be
C^1_1=bC^2_i
\ee
representing a single vector degree of freedom. Moreover,  one of the two $V_{i \alpha}$ is a pure gauge degree of freedom in the quasi-static approximation. This implies that two divergence-less vector degrees of freedom remain $C^1_1=bC^2_i$ and $V^i= V^i_1- V^i_2$.
Notice that the degree of freedom
$V^i= V^i_1- V^i_2$ is the one which leads in \cite{Comelli:2015pua} to a divergent mode. We will see that in the quasi-static approximation and at late times this mode is harmless.

It is now easy to see that the interaction term between the $C_{\alpha i}$ vanishes at the second order of perturbation theory as it would involve two time indices in the antisymmetric $\epsilon_{abcd}$ symbol.
{The same applies to the coupling between $C_{i\alpha}$ and $V^i$ which vanishes too.}  Hence
no contribution from the potential contains $C_{\alpha i}$ implying that these vectors have no extra potential terms beyond GR {at this order}. {On the other hand there are gradient terms in $(\partial _i V_j)(\partial^i V^j)$.}

Let us recall briefly how vectors behave in GR before generalising to the case of bigravity.
The quadratic Lagrangian in the quasi-static approximation is given by
\be
{\cal L}_V= -\frac{a^2}{32\pi G_N}(\vec \nabla C_i).(\vec \nabla C^i) - a^4 C^i \delta T^0_i
\ee
where we have used $\delta g_{0i} \delta T^{0i}=C^i \delta T_i^0$
{ and no gradient terms in $V_i$ appear as it can be formally gauged away. Moreover we focus on pressure-less matter which decouples from $\partial_i V_j$.}
The Euler-Lagrange equation becomes
\be
\Delta C_i=  16\pi G_N  a^2 \delta T^0_i.
\ee
Using $\delta T^0_i= (\rho+p) v_i$ where $v_i=\delta_{ij} v^j$ is the curl-part of the velocity fluid which decays like $1/a$, we find that $C_i$ decays like $1/a^2$ in the matter dominated era and can be neglected.

This can be generalised to the bigravity case where we use $C_i^2= b C_i^1$.
The Jordan frame vector field can be identified as
\be
a_J C^J_i= \beta_1 a_1 C^1_i+\beta_2 a_2 C^2_i
\ee
which implies that
\be
C^J_i=\frac{\beta_1 a_1 +\beta_2 b a_2}{a_J} C^1_i
\ee
while the Lagrangian for the bigravity vector field is
\be
{\cal L}_V= -\frac{a_1^2(1+b^2) }{32\pi G_N}(\vec \nabla C^1_i).(\vec \nabla C_1^i) - a_J^3 (\beta_1 a_1+\beta_2 b a_2)  C^i_J \delta T^0_{i}.
\ee
where $C^i_J= a_J^{-2} \delta^{ij} C^J_j$ implying that
\be
C^i_J= \frac{a_1^2}{a_J^2} \frac{\beta_1 a_1 +\beta_2 b a_2}{a_J} C^i_1.
\ee
The first term in the Lagrangian is the kinetic term coming from the two Einstein-Hilbert terms and the relation $C^i_2= b \frac{a_1^2}{a_2^2} C^i_2$ has been used.
We then deduce that
\be
\Delta C_i^1= 16\pi G_N \frac{(\beta_1 a_1 + \beta_2 a_2 b)^2}{1+b^2} \delta T^0_{i}
\ee
where $\delta T^0_{i}=(\rho+p) v_i$ implying that $C_i^1$ decays like $(\beta_1 a_1 + \beta_2 a_2 b)^2 a_J^{-4}$ in the matter era.
}{ The dynamics of $V_i$ simplify as the only terms in the Lagrangian involving $V_i$ are   gradient terms in $(\partial_i V_j)(\partial^i V^j)$ and no coupling to pressure-less matter appears, implying that
$V_i$ can be set to zero. Notice that $V_i$ behaves differently in the radiation era where a gradient instability is present, see section \ref{sec:insta}}.

{

\subsubsection{Tensor modes}

The gravitational sector is more interesting than the vector one. Focusing on the tensor perturbations
\be
\delta e^{\alpha i}_j= a_\alpha h^{i}_{\alpha j}
\ee
where $\alpha=1,2$ and $h^{i}_{\alpha j}$ is a symmetric transverse and traceless tensor with two degrees of freedom, we find that the mass term coming from the potential term of bigravity reads
\be
{\cal L}_m=12 m_{\alpha \beta} \Lambda^4 a_\alpha h^{i}_{j\alpha} a_\beta h^j_{i\beta}.
\ee
where
\be
m_{\alpha\beta}(a_\gamma)= \sum_{\gamma\delta} m_{\alpha\beta\gamma\delta}\tilde a_\gamma a_\delta.
\ee
where $\tilde a_\alpha= b_{\alpha}a_{\alpha}$ with $b_1=1$ and $b_2=b$.
The kinetic terms come from the two Einstein-Hilbert terms
\be
{\cal L}_L= \frac{1}{16\pi G_N}\left (a_1^2 \left ( \frac{dh_{ij}^1}{d\eta}\frac{dh^{ij}_1}{d\eta} -\vec\nabla h_{ij}^1\vec \nabla h^{ij}_1\right)+ \frac{a_2^2}{ b}\left (\frac{dh_{ij}^2}{d\eta}\frac{dh^{ij}_2}{d\eta} -b^2\vec\nabla h_{ij}^2\vec \nabla h^{ij}_2\right)\right ).
\ee
It is convenient to normalise the tensor modes according to
\be
 \bar h^1_{ij}=  m_{\rm Pl} a_1 h^1_{ij},\ \bar h^2_{ij}=  m_{\rm Pl} \frac{a_2}{b^{1/2}} h^2_{ij}
\ee
such that the kinetic terms become
\be
{\cal L}_L= \frac{1}{2}\left ( \frac{d\bar h_{ij}^1}{d\eta}\frac{d\bar h^{ij}_1}{d\eta} -\vec\nabla \bar h_{ij}^1\vec \nabla \bar h^{ij}_1+ \frac{1}{a_1} \frac{d^2 a_1}{d\eta^2} \bar h_1^{ij} \bar h^1_{ij}+ \frac{d\bar h_{ij}^2}{d\eta}\frac{d\bar h^{ij}_2}{d\eta} -b^2\vec\nabla \bar h_{ij}^2\vec \nabla \bar h^{ij}_2+ \frac{b^{1/2}}{a_2} \frac{d^2 (a_2b^{-1/2})}{d\eta^2} \bar h_2^{ij} \bar h^2_{ij}\right ).
\ee
The mass term becomes
\be
{\cal L}_m= 12 m^2 (b_\alpha b_\beta)^{1/2}m_{\alpha \beta}(a_\gamma) \bar h^{i}_{j\alpha}\bar h^j_{i\beta}.
\ee
and the mass matrix reads
\be
M^2_{\alpha\beta}(a_\gamma)= -24 m^2 (b_\alpha b_\beta)^{1/2}m_{\alpha\beta}(a_\gamma)
\label{mass}
\ee
which is a symmetric matrix of order $m^2$.

Let us consider first the Minkowski limit when $a_\alpha=1$. {In bigravity models, Minkowski space is not a solution of the Einstein equations as there is always a positive cosmological constant energy density
$ 24 \Lambda^2 \sum_{\alpha \beta} m_{\alpha \beta}$ at the background level. To obtain a model where Minkowski space is a solution of the equations of motions, we remove the contribution from the cosmological constant for the two metrics $g^\alpha_{\mu\nu}$, i.e. we consider the model with the action
\be
S\to S+ 24 \Lambda^4 (\int d^4x \sqrt{-g^1}   \sum_{ \beta} m_{1\beta}+ \int d^4x \sqrt{-g^2}  \sum_{\beta} m_{2\beta}).
\label{cv}
\ee
where $\delta g^{\alpha}_{ij}= 2 h^{\alpha}_{ij}$.
The corresponding Friedmann equations (\ref{F1}) and (\ref{F2}) have the solution $a_1=a_2=b=1$ associated to Minkowski space.
In this case, it is interesting to introduce the decomposition
\be
\bar h^i_{j1}= \frac{h^i_{j+} + h^i_{j-}}{\sqrt 2}, \ \ \bar h^i_{j2}=\frac{ h^i_{j+} - h^i_{j-}}{\sqrt 2}
\ee
and the change of basis induced by the matrix
\be
A=\left (
\begin{array}{cc}
1&1 \\
1&-1 \\
\end{array}
\right )
\ee
implying that, in the new basis, the mass matrix becomes
\be
\tilde M^2= AM^2A=-24 m^2 \left (
\begin{array}{cc}
m_{11}+2m_{12}+ m_{22}&m_{11}-m_{22} \\
m_{11}-m_{22}&m_{11}+m_{22}-2 m_{12} \\
\end{array}
\right ).
\ee
The Lagrangian from bigravity at the quadratic level becomes
\be
{\cal L}_T= -\frac{1}{2}( (\partial h^i_{+j})^2 +(\partial h^i_{j-})^2)- \frac{1}{2}\tilde M^2_{uv}h^{i}_{ju}h^j_{iv}
\ee
where $u,v=\pm$. When all the scale factors are equal to one, the background is consistent, i.e. Minkowski is indeed a solution as assumed above,  only when one removes the contribution to the mass of the gravitons coming from  the cosmological constants that we have introduced in (\ref{cv})
\be
{\cal L}_{cc}= 24 \Lambda^4 ( \sum_{ \beta} m_{1\beta} \sqrt{-g^1}+\sum_{\beta} m_{2\beta} \sqrt{-g^2}) \supset -\frac{1}{2} \Delta M^2_{uv} h^{i}_{ju}h^j_{iv}
\ee
where
the mass matrix coming from the added cosmological constant terms is
\be
\Delta M^2=24 m^2 \left (
\begin{array}{cc}
m_{11}+2m_{12}+ m_{22}&m_{11}-m_{22} \\
m_{11}-m_{22}&m_{11}+2m_{12}+m_{22} \\
\end{array}
\right )
\ee
leaving a total Lagrangian for the two gravitons $h_+$ and $h_-$
\be
{\cal L}_T= -\frac{1}{2}( (\partial h^i_{+j})^2 +(\partial h^i_{j-})^2)-\frac{1}{2}\bar M^2_{uv}h^{i}_{ju}h^j_{iv}
\ee
where we have introduced the mass matrix in a flat background
\be
\bar M^2= \tilde M^2 +\Delta M^2=  96 m^2 \left (
\begin{array}{cc}
0&0 \\
0& m_{12} \\
\end{array}
\right ).
\ee
Notice that the massless graviton is associated to $h_+$ (cf. the result of \cite{Schmidt-May:2014xla}) and the massive graviton to $h_-$ with a mass
\be
m^2_-= 96 m^2 m_{12}
\ee
which is always positive if we take the tensor $m_{abcd}$ to have only positive elements.
 It has to be emphasized that this mass matrix is not the mass matrix of a model of bigravity per se as we had to remove the cosmological constant terms in order to get a Minkowski background.

Let us come back to the case of a cosmological background.   The evolution equations for the two gravitons $\bar h_1$ and $\bar h_2$  are now
given by
\be
\frac{d^2{\bar {h }}_1}{d\eta^2}-\Delta \bar h_1 +(M^2_{11}(a_\gamma)- \frac{1}{a_1} \frac{d^2 a_1}{d\eta^2}) \bar h_1 + M^2_{12}(a_\gamma) \bar h_2=0
\label{pro1}
\ee
and
\be
\frac{d^2 {\bar {h }}_2}{d\eta^2}-b^2 \Delta \bar h_2 + ((M^2_{22}(a_\gamma)-\frac{b^{1/2}}{a_2} \frac{d^2 (a_2b^{-1/2})}{d\eta^2}) \bar {h_2} + M^2_{21}(a_\gamma) \bar h_1=0.
\label{pro2}
\ee
Notice that the two gravitons propagate at different speeds when $b\ne 1$.
Another new feature of bigravity is that the two gravitons $h_1$ and $h_2$ are coupled by the off-diagonal terms
of the mass matrix. This implies that there is gravitational birefringence and the two gravitons oscillate into one another as they propagate. This is analogous to what happens in the photon-axion or
photon-chameleon systems where birefringence implies a phase shift of the  waves. The analysis of these phenomenona is left for future work.

Let us finally comment on the coupling to matter. The Jordan frame matter couples to the combination
\be
a_J h^i_{jJ}= \beta_1 a_1 h^{i}_{j1}+ \beta_2 a_2 h^{i}_{j2}
\ee
and one can see that this evolves with time, i.e. matter couples to different gravitons in the history of the Universe. In the radiation and matter eras, the Jordan frame graviton simplifies to
\be
h^i_{jJ}= \frac{\beta_1^2 h^{i}_{j1}+ \beta_2^2  h^{i}_{j2}}{\beta_1^2 +\beta_2^2}
\ee
which differs from the Jordan frame graviton in the dark energy era.
}}

\section{Cosmological Evolution in Bigravity}
\subsection{The matter and radiation eras}

We only study the cosmological solutions of the model on the branch where
\be
b= \frac{a_2 H_2}{a_1 H_1}.
\ee
On this branch, the ratio $X= \frac{a_2}{a_1}$ is algebraically determined. In particular, the influence of the potential term of bigravity, as we have taken the mass term $m\sim H_0$, only plays a role on the background cosmology in the late time Universe. This is very particular to this branch of solutions and this would not be the case on the other branch  where the pressure and dark energy are directly related.
In the early Universe and on the branch (\ref{bi}), i.e. in the radiation and matter eras along this branch, we will neglect the potential term and study the evolution of the Universe due to the double coupling to matter. We already know that in this regime we have that $X=a_2/a_1 \to \beta_2/\beta_1$. We will go into more details of the dynamics of the model in the matter-radiation eras.

In the matter-radiation eras the matter term in $\rho$ dominates over the potential term in $\Lambda^4$, this implies that the Friedmann equations read
\be
3H_1^2 m_{\rm Pl}^2= \beta_1 \frac{a_J^3}{a_1^3}\rho
\ee
and
\be
\frac{3H_2^2 m_{\rm Pl}^2}{b^2}= \beta_2 \frac{a_J^3}{a_2^3}\rho .
\ee
A family of solution can be obtained when the two scale factors are proportional
\be
a_2= X a_1
\ee
implying that the $b$ factor is also constant
as we have
\be
3H_1^2 m_{\rm Pl}^2= \beta_1 (\beta_1 +\beta_2 X)^3 \rho
\ee
and
\be
\frac{3H_2^2 m_{\rm Pl}^2}{b^2}= \frac{3H_1^2 m_{\rm Pl}^2}{X^2b^2}= \beta_2 \frac{(\beta_1 +\beta_2 X)^3}{X^3}\rho,
\ee
from which we deduce that
\be
b^2=  \frac{\beta_1}{\beta_2}X.
\ee
The Raychaudhuri equations become
\be
2m_{\rm Pl}^2  \frac{1}{a_1^3} \frac{d^2 a_1}{d\eta^2}=m^2_{\rm Pl} H_1^2 - \beta_1 \frac{e}{e_1} \frac{\beta_1 a_1^2 + \beta_2 a_1 a_2 }{a_J^2} p
\ee
and similarly
\be
2m_{\rm Pl}^2  \frac{1}{b^2a_2^3} \frac{d^2 a_2}{d\eta^2}=m^2_{\rm Pl} \frac{H_2^2}{b^2} - \beta_2 \frac{e}{e_2} \frac{\beta_2 a_2^2 + \beta_1 a_1 a_2 }{a_J^2} p.
\ee
This becomes
\be
2m_{\rm Pl}^2  \frac{1}{b^2X^2a_1^3} \frac{d^2 a_1}{d\eta^2}=m^2_{\rm Pl} \frac{H_1^2}{b^2X^2} - \frac{\beta_2}{bX^4} \frac{e}{e_1} \frac{a_1^2 (\beta_2 X^2 + \beta_1 X) }{a_J^2} p
\ee
which implies that
\be
b=\frac{\beta_1}{\beta_2} X.
\ee
We then deduce that the ratio of the lapse functions must be equal to one, i.e.
\be
b=1, \ X=X_m=\frac{\beta_2}{\beta_1}.
\ee
Let us confirm  that the Raychaudhuri equation is consistent with this solution.
The conservation of matter leads to
\be
\rho= \frac{\rho_0}{a_J^{3(1+\omega)}}=\frac{\rho_1}{a_1^3}
\ee
where $\rho_1=\rho_0 (\beta_1+\beta_2 X)^{3(1+\omega)}$ is a constant.
Defining the cosmic time $dt_1= a_1 d\eta$, we have the following time evolution for the scale factor
\be
a_1=(\frac{3}{2} (1+\omega) \frac{t_1}{t_K})^{2/3(1+\omega)}
\ee
where we have defined the characteristic time $t_K^{-1}= \frac{\beta_1\rho_1}{3m_{\rm Pl}^2} \frac{e}{e_1}  \frac{1}{(\beta_1 +\beta_2 bX)}$ as a constant. Using
\be
\frac{d^2a_1}{a_1^3 d\eta^2}= \frac{d^2 a_1}{a_1dt_1^2} +H_1^2
\ee
we find the equality between the constants of the model
\be
1- \frac{3(1+\omega)}{2} = -\frac{1}{2}-3\omega \frac{K_1}{K_2}
\ee
where
the coefficients are
\be
K_{1}=\frac{\beta_1+ \beta_2 X b}{(\beta_1 +\beta_2 bX)^2},\ \ K_2= \frac{\beta_1 + \beta_2 X  }{(\beta_1+ \beta_2  X)^2}.
\ee
This implies that these
 constants  must be equal
\be
K_1=K_2
\ee
and finally we find the same conditions
\be
b=1,\ X=X_m=\frac{\beta_2}{\beta_1}.
\ee
With this we have that the dynamics of the Universe in the matter-radiation eras are determined by
\be
H_1^2= \beta_1 (\beta_1 + \beta_2 X)^3 \frac{\rho}{3m_{\rm Pl}^2}
\ee
and
\be
H_2^2=\beta_2 \frac{(\beta_1+ X\beta_2)^3}{X^3}  \frac{\rho}{3m_{\rm Pl}^2}
\ee
which coincides with $H^2_2=H_1^2/X^2$.
As a result the Hubble rate in the Jordan frame is given by $H_J=\frac{H_1}{\beta_1+\beta_2 X}$ and the Friedmann equation  reads
\be
H_J^2 =(\beta_1^2 +\beta_2^2)\frac{\rho}{3m_{\rm Pl}^2}.
\ee
This confirms that the dynamics on the branch (\ref{bi}) in the matter-radiation eras follow a Friedmann equation like in GR. The only big difference is that
the Friedmann equation depends on  the background Newton constant in the matter and radiation eras
\be
\frac{G_{N \rm{cosmo}}}{G_N}= \beta_1^2+ \beta_2^2
\ee
which needs to be compared to local tests of gravity (see below). If $G_{N\rm{cosmo}}\ne G_{N}^{\rm local}$ then the background cosmology in the matter-radiation eras would differ from the $\Lambda$-CDM dynamics which satisfies
\be
H_{\Lambda CDM}^2 =8\pi G_N^{\rm local}\frac{\rho}{3}.
\ee
We will analyse the link between $G_{N{\rm cosmo}}$ and $G_N^{\rm local}$ below and we will in fact find that they coincide implying that the matter-radiation eras along the branch (\ref{bi}) and in the $\Lambda$-CDM model agree.
We also have that in these eras the slip parameter is given by
\be
\eta=1
\ee
as $b=1$ and
\be
\mu=(\beta_1^2 +\beta_2^2)\frac{G_N}{G_N^{\rm local}}.
\ee
We will calculate $\mu$ using local experiments in the next section, i.e. after determining $G_N^{\rm local}$.

As the matter density $\rho \sim a_1^{-3 }$ decreases in the matter era, the contribution from the potential term of massive bigravity becomes less subdominant.
Notice that the potential term contributes a constant term to the Friedmann equations for $H_{1,2}$
\be
\Lambda_1^4= 24\Lambda^4 m^{1jkl}\frac{a_j a_k a_l}{a_1^3}
\ee
and
\be
\Lambda_2^4= 24\Lambda^4 m^{2jkl}\frac{a_j a_k a_l}{a_2^3}
\ee
which act as subdominant cosmological constants in the radiation and matter eras.
When these terms start to dominate, bigravity acts as dark energy.
\begin{figure*}
\centering
\includegraphics[width=0.49\linewidth]{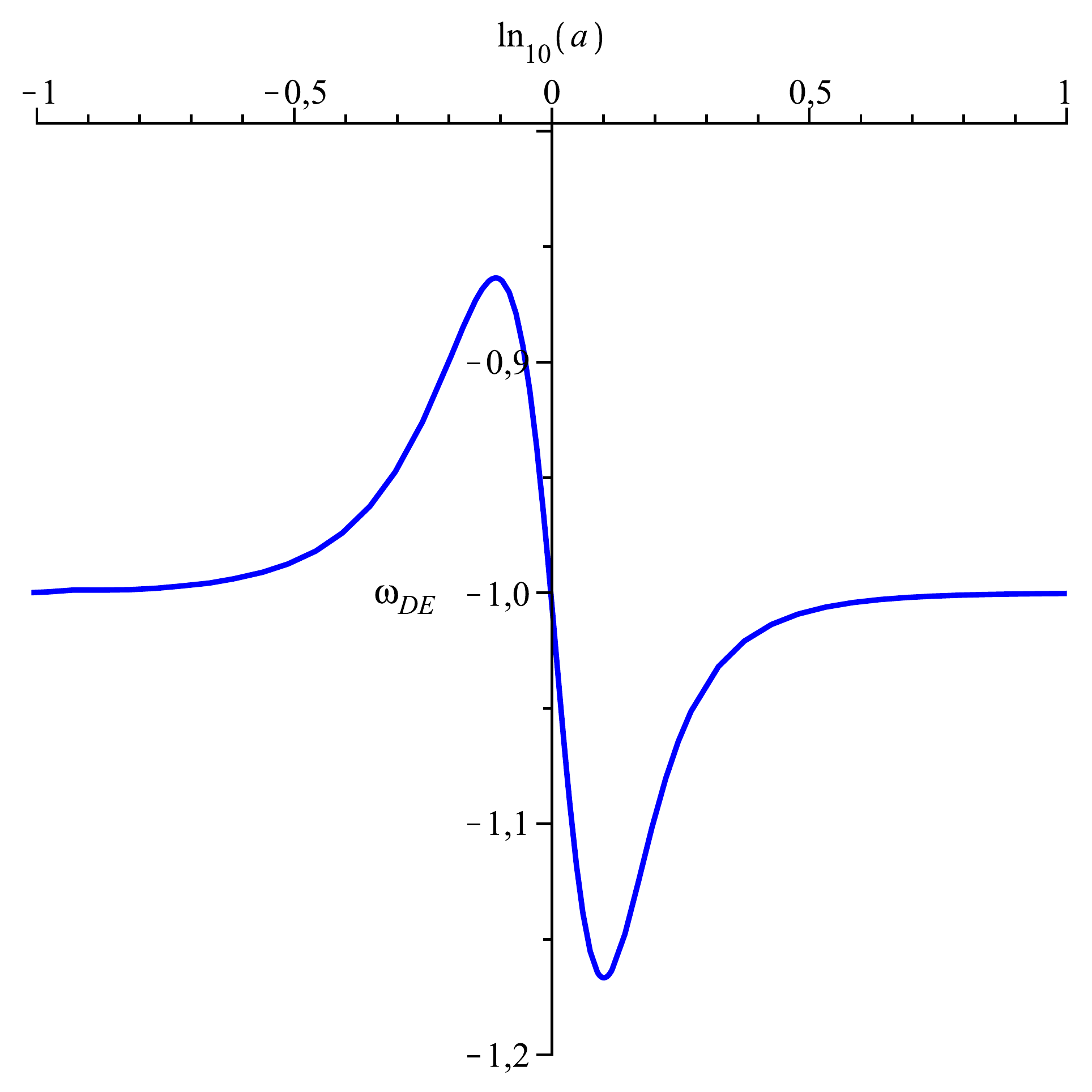}
\includegraphics[width=0.49\linewidth]{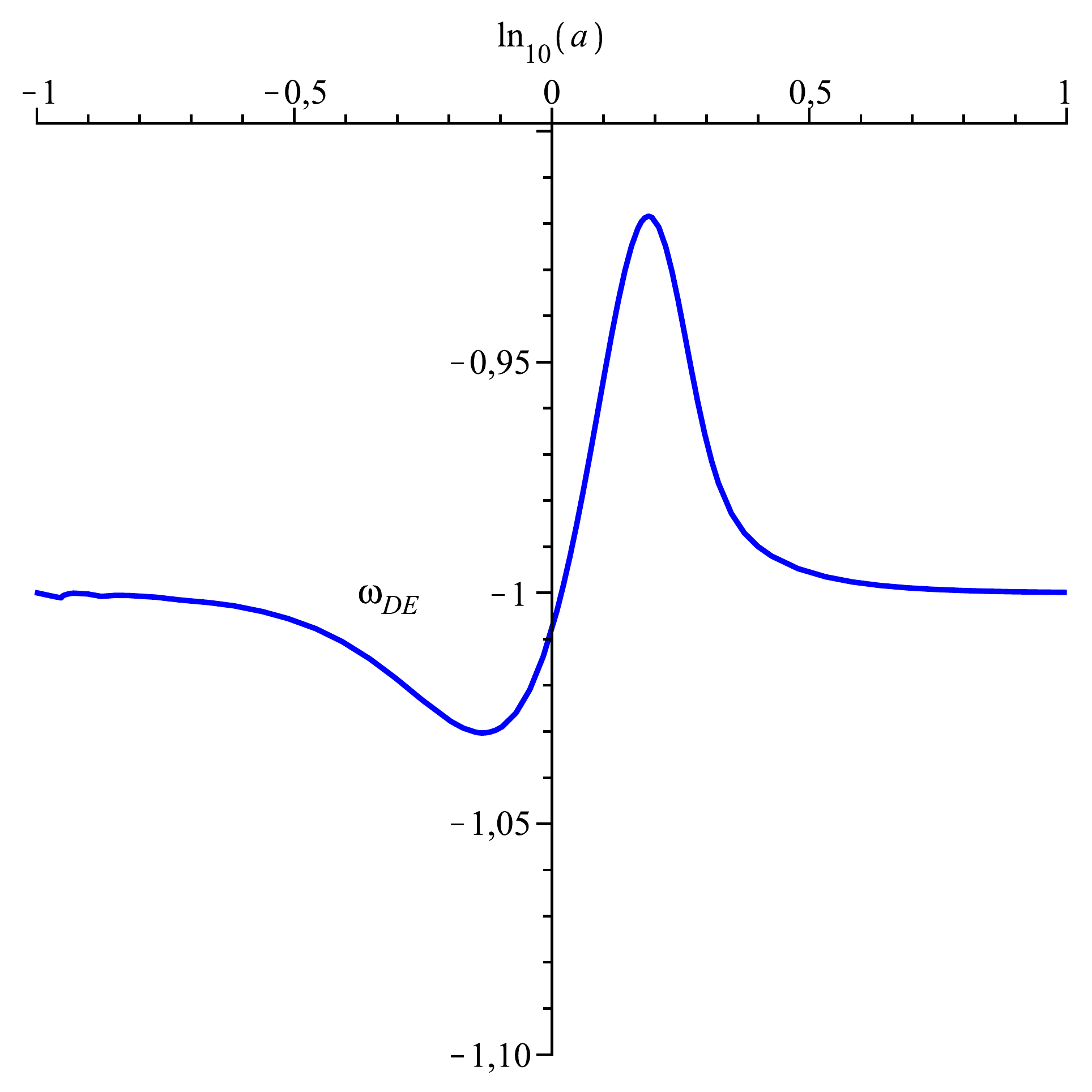}
\caption{The variation of the equation of state $w_{\rm DE}$  as a function of  redshift $a_J$ from  $a_{\rm ini}=10^{-4}$ with model I (left panel) and model II (right panel). }
\end{figure*}

\begin{figure*}
\centering
\includegraphics[width=0.49\linewidth]{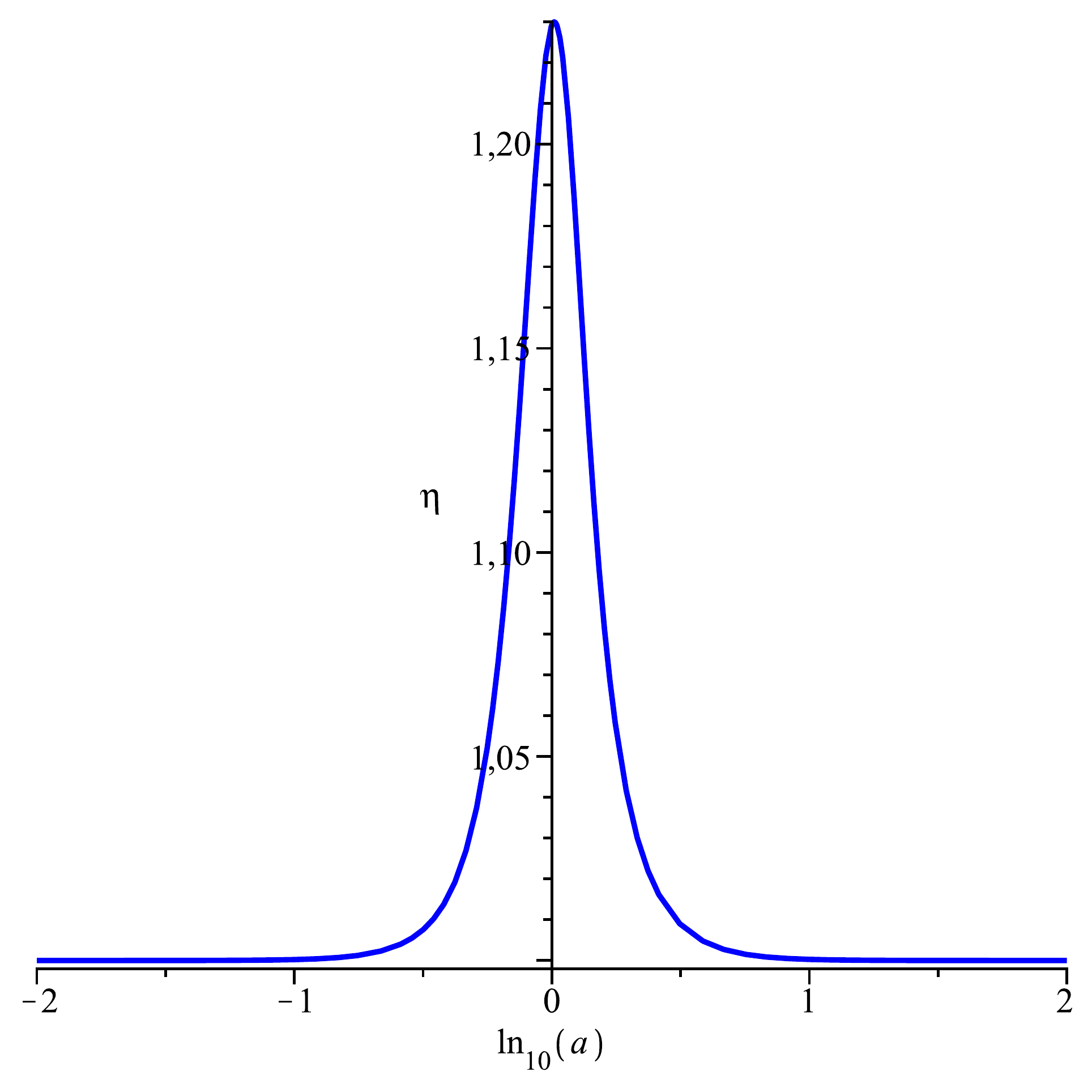}
\includegraphics[width=0.49\linewidth]{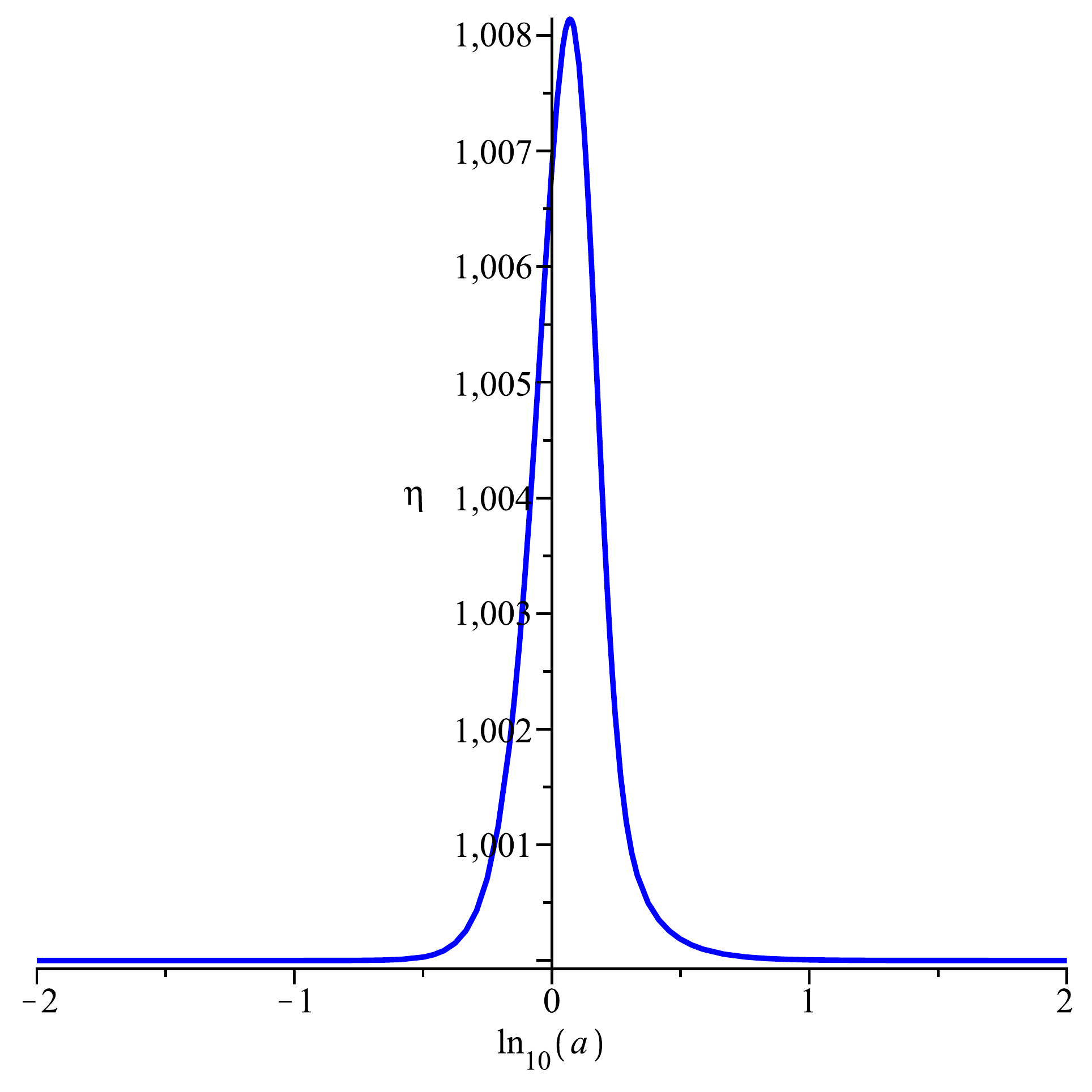}
\caption{The variation of slip function  $\eta$  as a function of  redshift $a_J$ from  $a_{\rm ini}=10^{-4}$ with model I (left panel) and model II(right panel). }
\end{figure*}

\subsection{Dark energy}

When the matter density becomes subdominant, the Friedmann equations reduce to
\be
3H_1^2 m_{\rm Pl}^2= 24\Lambda^4 m^{1jkl}\frac{a_j a_k a_l}{a_1^3}
\ee
and
\be
\frac{3H_2^2 m_{\rm Pl}^2}{b^2}= 24\Lambda^4 m^{2jkl}\frac{a_j a_k a_l}{a_2^3}.
\ee
The Hubble rates become constant and the space-time becomes de Sitter with
\be
a_2=X a_1
\ee
where we must have
\be
b^2 =X \frac{m^{1jkl}{a_j a_k a_l}}{m^{2jkl}{a_j a_k a_l}}.
\ee
Using the Raychaudhuri equation
\be
2m_{\rm Pl}^2  \frac{1}{b^2a_2^3} \frac{d^2 a_2}{d\eta^2}=m^2_{\rm Pl} \frac{H_2^2}{b^2} +  24 \Lambda^4 m^{2jkl}\frac{\tilde a_j a_k a_l}{ba_2^3}
\ee
and $\frac{d^2 a_1}{a_1^3 d\eta^2}= 2H_1^2$ we find that
\be
b=\frac{m^{2jkl}{\tilde a_j a_k a_l}}{m^{2jkl}{a_j a_k a_l}}
\ee
whose solution is
\be
b=1.
\ee
Therefore we find that $X\to X_d$ where
\be
X_d=\frac{m^{2jkl}{a_j a_k a_l}}{m^{1jkl}{a_j a_k a_l}}
\ee
which can be written explicitly as
\be
X_d= \frac{m^{2222}X_d^3 + 3 m^{2221} X_d^2 + 3 m^{2211} X_d + m^{2111}}{m^{1111} + 3 m^{2111} X_d + 3 m^{2211} X_d^2 + m^{1222} X_d^3}.
\label{xd}
\ee
 Only models with positive real roots admit a late time dark energy era. This depends on the choice of the couplings $m^{ijkl}$. When the above equation admits no solution, the ansatz $a_2=X a_1$ does not lead to
 meaningful solutions anymore and more complex solutions must be looked for.

In this dark energy phase, if existent,  the Newtonian potentials satisfy the same properties as in the matter and radiation eras
\be
\eta=1, \ \ \mu= (\beta_1^2 +\beta_2^2)\frac{G_N}{G_N^{\rm local}}
\ee
where the growth of structure depends on the value of $G_N^{\rm local}$.

{
\subsection{Instabilities}
\label{sec:insta}

We can now discuss the issue of gravitational and vector instabilities and the validity of the quasi-static approximation. First of all we have seen that the mass matrix of the gravitons (\ref{mass}) has only negative entries as long as $m_{ijkl}\ge 0$. The positivity of the coefficients $m_{ijkl}$ guarantees that at all times the effective cosmological constant provided by the potential term of bigravity is positive, i.e. this evades possible big crunch singularities if the  potential became negative. A negative mass matrix signals potential tachyonic instabilities. This can be analysed using the propagation equations (\ref{pro1}) and (\ref{pro2}).
In the matter dominated era where $X$ is constant and $b=1$, the mass matrix  is dominated by the  diagonal terms $\frac{1}{a_\alpha} \frac{d^2 a_\alpha}{d\eta^2}$ of  order ${\cal H}^2_{1,2}$ respectively and only modes $\bar h_{\alpha}$ such that $k/a_{1,2} \lesssim  H_{1,2}$, i.e. modes outside the horizon, grow in $a_\alpha$ implying that $ h_{\alpha}$ remains constant. Hence there is no instability in the matter era. In the radiation dominated era where $b=1$ and $X$ is constant too, the diagonal terms $\frac{1}{a_\alpha} \frac{d^2 a_\alpha}{d\eta^2}$ vanish.  There are now new   pressure-dependent mass terms coming from the coupling to matter which read
\be
\delta S_p=  \frac{1}{8} \int d^4x e \delta T_{ij}\delta g^{ij}
\ee
where $\delta T_{ij}= 2 a_J( \beta_1 a_1 h_{ij}^1 +\beta_2 a_2 h^22_{ij}) p$ and $\delta g^{ij}= -2 a_J^{-3} (\beta_1 a_1 h_{ij}^1 +\beta_2 a_2 h^22_{ij})$.
There is also a term coming from the two Einstein-Hilbert contributions
\be
\delta S_g= -\frac{1}{16\pi G_N}\int d^4 x \left ( e_1 (2 \frac{dH_1}{dt_1} + 3 H_1^2) h_1^{ij}h^1_{ij}+ e_2 (2 \frac{dH_2}{dt_2} + 3 H_2^2) h_2^{ij}h^2_{ij}\right).
\ee
Using $2 \frac{dH_\alpha}{dt_\alpha} + 3 H_\alpha^2= -3\omega H_\alpha^2$ where $\omega=1/3$ and $a_\alpha H_\alpha= a_J H_J$, we find that
\be
\delta S_g= \int d^4x \frac{3\omega a_J^2 H_J^2}{2}(\bar h_1^{ij}\bar h^1_{ij}+
 \bar h_2^{ij}\bar h^2_{ij})
\ee
and the matter term
\be
\delta S_p= -\int d^4x \frac{3\omega a_J^2 H_J^2}{2(\beta_1^2+\beta_2^2)}(\beta_1\bar h_1^{ij}+\beta_2
 \bar h_2^{ij})^2.
\ee
As a result we find that in the radiation dominated era, the pressure mass matrix becomes
\be
 \Delta  M^2_p= \frac{3\omega a_J^2 H_J^2}{\beta_1^2+\beta_2^2}  \left (
\begin{array}{cc}
-\beta_2^2&\beta_1\beta_2  \\
\beta_1\beta_2 & -\beta_1^2\\
\end{array}
\right ).
\ee
Deep in the radiation era, the correction term $\Delta  M_p^2$ dominates. Notice that the pressure dependent matrix has always a zero mass eigenstate (in practice the mass of this eigenstate comes from the neglected terms and is very small compared to the Hubble rate) and an eigenmode of negative mass
\be
m^2_G=-3\omega a_J^2 H_J^2<0
\ee
corresponding to an instability for modes outside the cosmological horizon. This instability has a growing factor $D_+$ which satisfies
\be
D_+'' -\frac{1}{\eta^2}=0
\ee
which grows like
\be
D_+\sim a^{\lambda_+},\ \ \lambda_+= \frac{1+\sqrt{5}}{2}.
\ee
The normalised zero eigenmode is given by
\be
h^+_{ij}= \frac{\beta_1 \bar h^1_{ij} + \beta_2 \bar h^2_{ij}}{\sqrt{\beta_1^2 +\beta_2^2}}= \frac{a_1h^J_{ij}}{\sqrt{8\pi G_{\rm N cosmo}}}
\ee
corresponding to the Jordan frame graviton normalised by the cosmological Newton constant.
The massive eigenmode is
\be
h^-_{ij}= \frac{\beta_2 \bar h^1_{ij} - \beta_1 \bar h^2_{ij}}{\sqrt{\beta_1^2 +\beta_2^2}}=\frac{\beta_2}{\sqrt{\beta^2_1+\beta_2^2}} m_{\rm Pl} a_1 (h^1_{ij}-h^2_{ij})
\ee
implying a mild growth of the gravitons $h^\alpha_{ij}$ in $a^{\frac{\sqrt{5}-1}{2}}$ outside the horizon \cite{Comelli:2015pua}.

The same reasoning can be applied to the two vectors $V_{i\alpha}$ beyond the quasi-static approximation. Defining
\be
 \bar V^1_{ij}=  m_{\rm Pl} a_1 V^1_{i},\ \bar V^2_{i}=  m_{\rm Pl} a_2 V^2_{i},
\ee
the gradient terms read
\be
{\cal L}_V=-(M^2 +\Delta M_p^2)_{\alpha\beta}(\partial_i \bar V_j^\alpha) (\partial^i \bar V^{j\beta})
\ee
which shows a gradient instability when the tensor mass matrix has negative eigenvalues \cite{Gumrukcuoglu:2015nua}. This is the case in the radiation dominated era where the pressure mass term dominates. As for the tensor perturbations, the
Jordan frame vector
\be
a_J V_J^i = \beta_1 a_1 V_1^i+\beta_2 a_2 V_2^i
\ee
corresponds to the zero eigenmode with no gradient instability. On the contrary, the mode
\be
V^-_{i}= \frac{\beta_2 \bar V^1_{i} - \beta_1 \bar V^1_{i}}{\sqrt{\beta_1^2 +\beta_2^2}}=\frac{\beta_2}{\sqrt{\beta^2_1+\beta_2^2}} m_{\rm Pl} a_1 (V^1_{i}-V^2_{i})
\ee
is the unstable mode with a gradient instability.
In conclusion, we have retrieved the fact that vectors and tensors can be unstable in the radiation dominated era \cite{Comelli:2015pua}. The Jordan frame vector and tensor perturbations, i.e. the ones which couple to matter, do not suffer from such instabilities. Eventually, it  would remain to be seen how lethal these instabilities in sectors decoupled from matter are.

{Finally we would like to re-emphasise that}, in this paper, we consider bigravity theories at low energy, i.e. from the late radiation era to the dark energy one. Indeed at higher energies the UV completion of bigravity {most likely} would modify the behaviour of the theory and possibly alter either the presence or the type  of instabilities. At low energy, i.e. where we are safely in the regime of validity of the theory and can most trust it, no instability is present and all the mass matrices for the various perturbations which come from the potential term of bigravity are negligible compared to the large gradient terms in the sub-horizon limit. As a result, in the sub-horizon limit and at low energy we can use the quasi-static approximation for the time evolution of perturbations as shown in previous sections. }

\section{Local Dynamics}
\label{sec:local}
\subsection{Local gravity}

We are interested in gravity tests performed in the solar system. In these cases, the Newtonian potential is very small hence the background geometry is well approximated by Minkowski space-time. This would not be the case around neutron stars for instance where another treatment is required. Following our analysis of the scalar degrees of freedom, we know that there are four Newtonian potentials $(\Psi_{1,2},\Phi_{1,2})$. {In the quasi-static approximation and as long as the Newtonian potentials are small, e.g. in the solar system, the fifth degree of freedom decouples from matter and the Newtonian potentials.}
In such a Minkowski background we consider an over-density of matter determined by the matter density $\delta \rho$. The full Lagrangian of the gravitational dynamics including the four potential terms
up to second order is simply
\bea
&&
{\cal L}= \frac{1}{8\pi G_N} ((\vec\nabla \Psi_1)^2-2 \vec\nabla\Psi_1 . \vec\nabla \Phi_1)+ \frac{1}{8\pi G_N} ((\vec\nabla \Psi_2)^2-2 \vec\nabla\Psi_2 . \vec\nabla \Phi_2)
- \delta\rho ({\beta_1 \Phi_1+ \beta_2 \Phi_2})({\beta_1+\beta_2})^3\nonumber \\ && -72\Lambda^4 m^{ij} (\Psi_i-\Phi_i) \Psi_j\nonumber \\
\eea
where $m^{ij}=\sum_{kl}m^{ijkl}$.
From this we deduce the Poisson equations
\be
\Delta \Psi _i = \beta_i (\beta_1+\beta_2)^3 4\pi G_N\rho  -72 \times 4\pi G_N \Lambda^4 m^{ij} \Psi_j
\ee
and
\be
\Delta \Phi _i = \beta_i (\beta_1+\beta_2)^3 4\pi G_N\rho  -72 \times 4\pi G_N \Lambda^4 m^{ij}(\Phi_j- \Psi_j).
\ee
We focus on distances much less that $1/m$ implying that one can safely neglect the mass terms and get the two Poisson equations
\be
\Delta_{\rm phys} \Phi_J= 4\pi G_N (\beta_1^2 +\beta_2^2)\delta  \rho
\ee
and
\be
\Delta_{\rm phys} \Psi_J= 4\pi G_N (\beta_1^2 +\beta_2^2)\delta \rho
\ee
from which we find that the local Newtonian potential is $\Phi_N=\Psi_J=\Phi_J$. Doing so, we have defined the physical coordinates as $(\beta_1+\beta_2) \vec x$. We can now identify the Newton constant $G_N(\beta_1^2 +\beta_2^2)$  with the one measured locally
\be
G_N^{\rm local}= (\beta_1^2 +\beta_2^2) G_N.
\ee
This implies that the equality between  the local and background cosmological values of Newton's constant is satisfied
\be
G_N^{\rm local}=G_{N\rm cosmo}.
\ee
As a result we have that in the matter, radiation and dark energy eras
\be
\eta=1, \ \ \mu=1
\ee
with no modification of gravity.

{Notice that the local dynamics do not require the presence of a Vainshtein mechanism to screen the existence of a propagating massless scalar. The only scalar on top of the Newtonian potentials
is the $U$ field which decouples from matter. This is analogous to the absence of Vainshtein mechanism in massive dRGT gravity with a single coupling in the decoupling limit \cite{deRham:2010tw}. Here we find it at the bigravity level in the doubly coupled case.

In fact there {appears to be} a fundamental reason why the Vainshtein mechanism is not necessary in the doubly coupled case. Indeed when a single matter species is coupled in bigravity, the matter action does not break the two copies of diffeomorphism invariance of the theory. This implies that in the low energy limit, i.e. {the $\Lambda_3$ decoupling limit where $\Lambda_3=m^2 m_{\rm Pl}$, is kept fixed and matter fields are scaled such that their lowest energy contribution is kept in the Lagrangian}, the \St{} field does not couple to matter before demixing with gravity. The demixing introduces a direct, i.e. {linear and non-derivative}, coupling of the \St{} field to matter, which then needs to be Vainshtein-screened in a Galileon fashion. In the doubly coupled case, the \St{} field is {already present in the matter coupling prior to demixing, due to the diffeomorphism breaking nature of the matter coupling.
Consequently, the lowest energy contribution from the matter coupling now immediately comes in at the $\Lambda_3$ level , i.e. no further scaling of the matter content is required,  and introduces a direct derivative coupling between the scalar and matter. We will discuss this in detail in \cite{us}. Notice that, when taking the decoupling limit at the $\Lambda_3$ scale without scaling matter, in this limit derivative interactions with pressure-less matter vanish in the static limit as  only the $T_{00}$ component of matter contributes and time derivatives of the \St{} field vanish. This precludes the necessity for the Vainshtein mechanism in this limit. However, the same non-derivative couplings to matter as in the singly coupled are still present  at higher energy scales, so some amount of Vainshtein screening beyond the static and decoupling limits will be required and present.}}

\subsection{Local tests}

As the Poisson equations are not modified in a Minkowski background, the orbits of planets are not affected. The only local deviation from Newtonian gravity follows from the
slight time dependence of the Newton constant as the geometry is locally FRW and influenced by the background cosmology. As the Poisson equations are linear, we can
superimpose the solutions for all the objects in the Milky Way as embedded in the  cosmological background. This implies that the planetary orbits depend on
\be
\Delta \Phi_J= 4\pi G_N^{\rm local}\mu \delta \rho
\ee
where $\Delta$ is the Laplacian in the physical coordinates. In particular the Lunar Ranging experiment which triggers the motion of the moon in the solar system implies that a time drift of Newton's constant is severely constrained \cite{Williams:2004qba}
\be
\vert \frac{d\ln G_N^\Phi}{dt_J}\vert = \vert \frac{d\ln \mu}{dt_J}\vert \le 0.02 H_0
\label{Gdot}
\ee
at the present time.
We have seen that $\mu=1$ in the matter and dark energy eras. This implies that $\mu$  can only vary in the transient region when $b\ne 1$ and $X$ evolves between its matter dominated value $X_m$  to its dark energy value $X_d$.

\begin{figure*}
\centering
\includegraphics[width=0.49\linewidth]{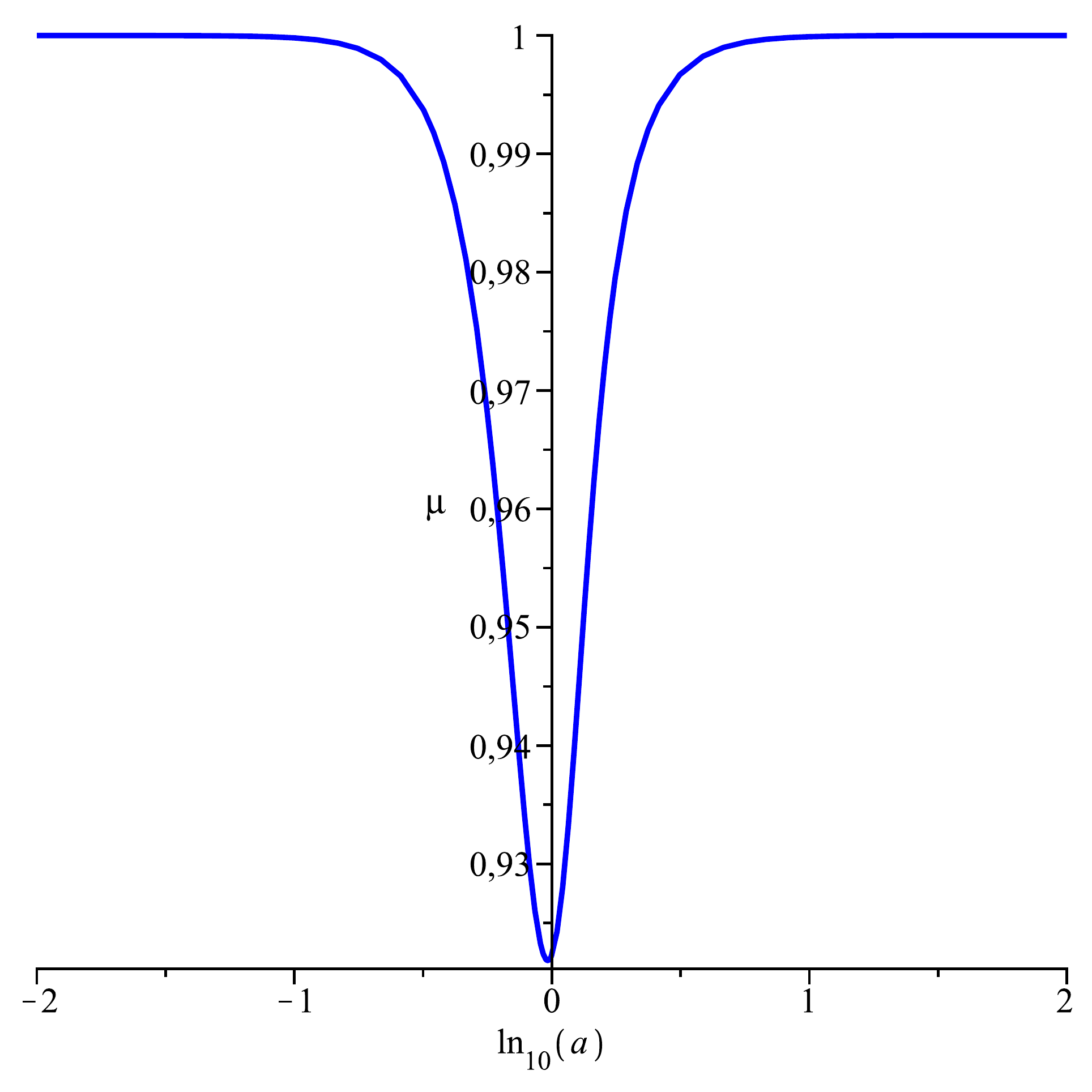}
\includegraphics[width=0.49\linewidth]{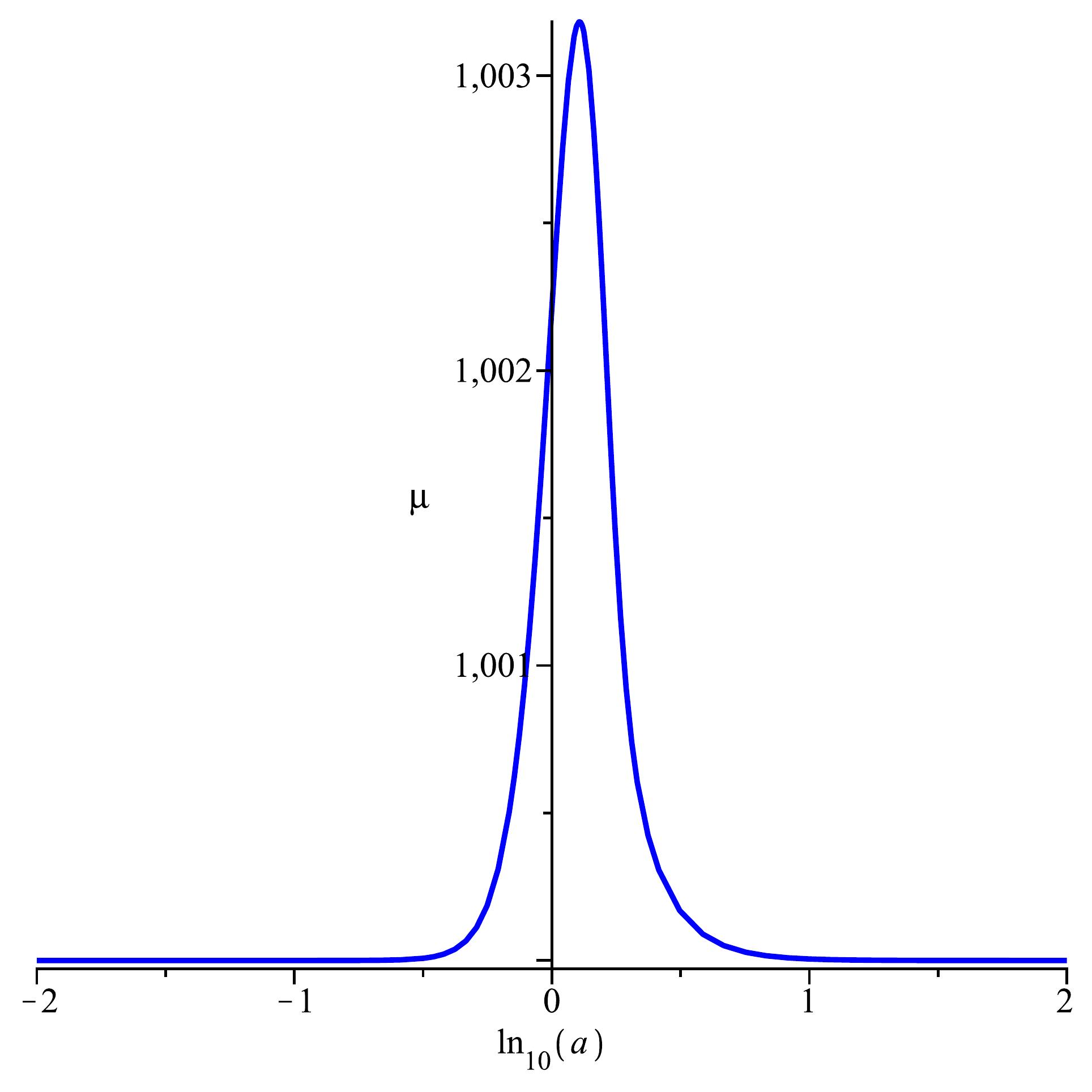}
\caption{The variation of  the growth parameter $\mu$  as a function of  redshift $a_J$ from  $a_{\rm ini}=10^{-4}$ with model I (left panel) and  model II (right panel).}
\end{figure*}

\begin{figure*}
\centering
\includegraphics[width=0.49\linewidth]{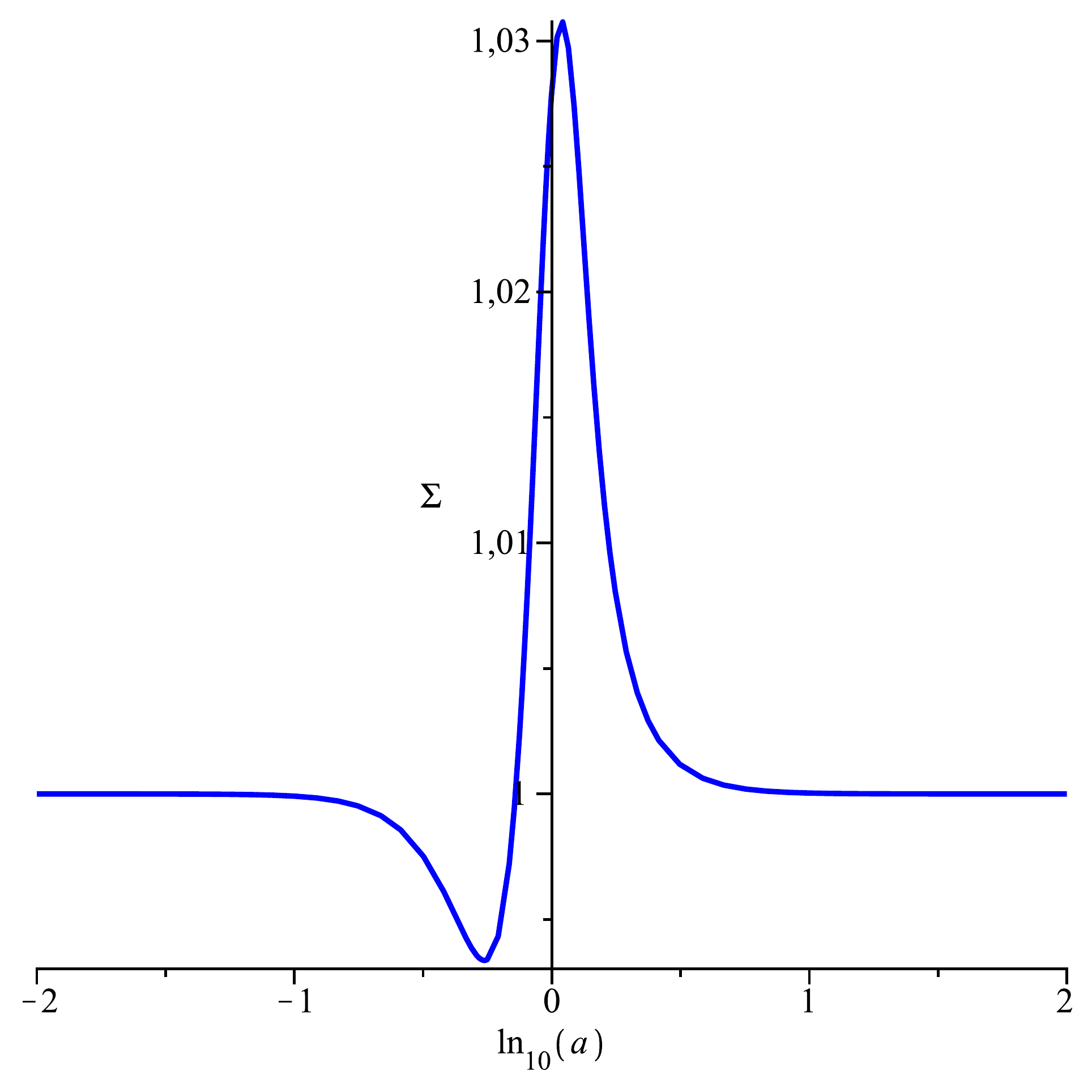}
\includegraphics[width=0.49\linewidth]{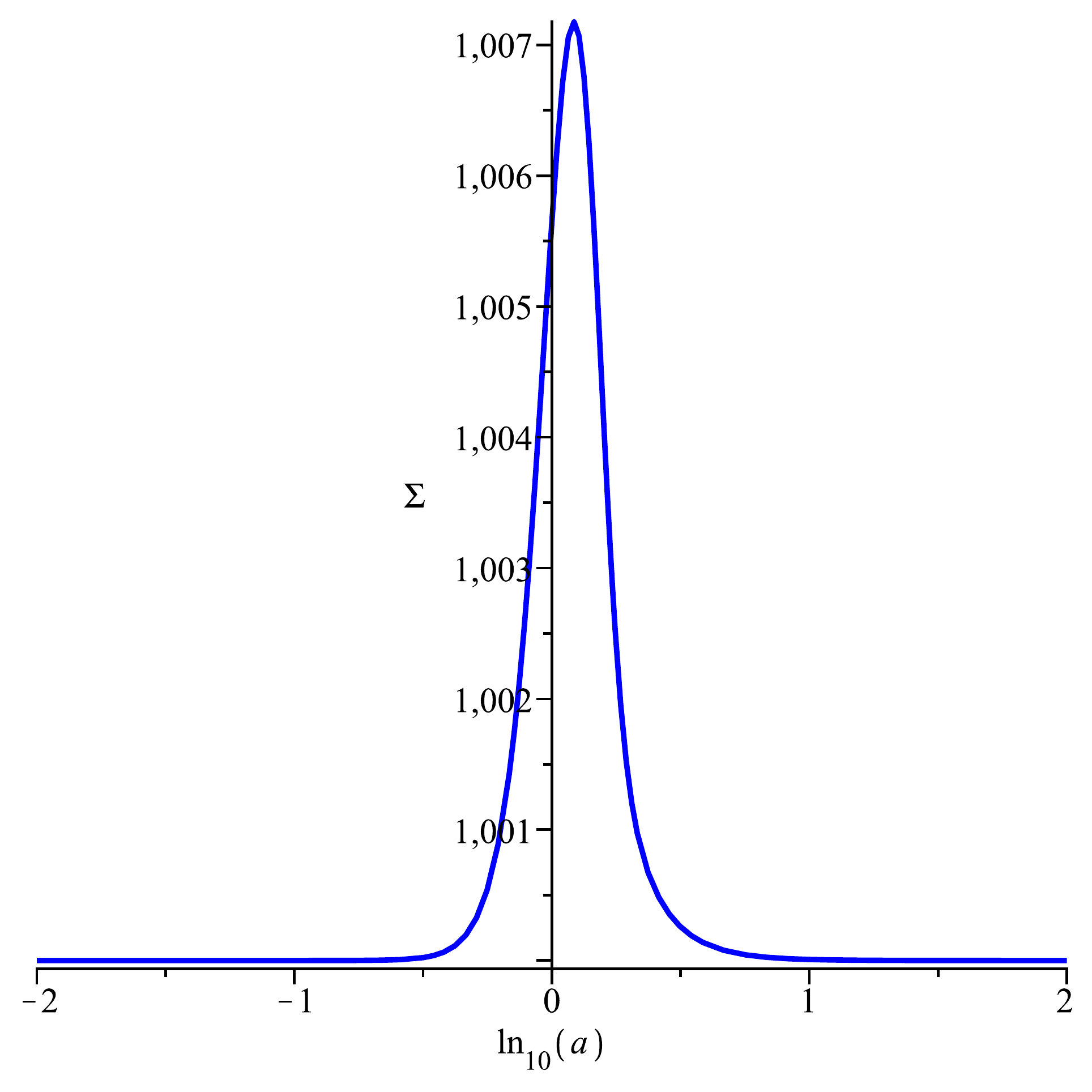}
\caption{The variation of  $\Sigma$  as a function of  redshift $a_J$ from  $a_{\rm ini}=10^{-4}$ with model I (left panel) and model II (right panel). }
\end{figure*}

\section{Numerical Results}

\subsection{Cosmological Evolution and Modified gravity}

We focus on the branch of solutions where $b=\frac{a_2 H_2}{a_1H_1}$ only. In this case, the matter and late radiation eras are retrieved. Moreover the modification of gravity that could be  induced on the growth of structure and lensing is absent  on cosmological scales as $\mu=\eta=\Sigma=1$. Similarly when $X_d$ exists as a solution of (\ref{xd}), i.e   in the dark energy era, gravity is not modified too. Hence gravity can only be modified with an impact on $\eta$, $\mu$ and $\Sigma$ in the intermediate regions where $X$ goes from its matter-radiation value $X_m$ to its dark energy one $X_d$. During this transition, if $b\ne 1 $, then $\eta\ne 1$ and $\mu\ne 1 $. Modified gravity then appears  only as  a transient phenomenon which would be taking place at the present epoch in the history of the Universe.

The cosmological evolution can be numerically analysed using the number of e-folds
\be
N=\ln a_J
\ee
in the Jordan frame.
The dynamics reduce to a system of three first order differential equations for $\ln b$ and $\ln a_{1,2}$. We have first
\be
\frac{d\ln a_1}{dN}= \frac{\beta_1 + \beta_2 X}{\beta_1 + \beta_2 X b}, \ \ \ \frac{d\ln a_2}{dN}= b \frac{\beta_1 + \beta_2 X}{\beta_1 + \beta_2 X b}
\ee
where we have defined the reduced Hubble rates
\be
\bar H_{1,2}= \frac{H_{1,2}}{H_0}
\ee
and we normalise
\be
\rho= \frac{\rho_0}{a_J^3}
\ee
where $\rho_0= \frac{3\Omega_m^{(0)} H_0^2}{8\pi G_N^{\rm local}}= \frac{3\Omega_m^{(0)} H_0^2 m_{\rm Pl}^2}{\beta_1^2 +\beta_2^2}$. The reduced Hubble rates are therefore
\be
\bar H_1^2 =  \frac{\beta_1}{ \beta_1^2+\beta_2^2}\frac{\Omega_m^{(0)}}{a_1^3} +8\frac{\Lambda^4}{m_{\rm Pl}^2H_0^2} m^{1jkl}\frac{a_j a_k a_l}{a_1^3}
\ee
Similarly we find that
\be
\frac{\bar H_2^2 }{b^2}=  \frac{\beta_2}{\beta_1^2+\beta_2^2} \frac{\Omega_m^{(0)}}{a_2^3} +8\frac{\Lambda^4}{m_{\rm Pl^2}H_0^2} m^{2jkl}\frac{a_j a_k a_l}{a_2^3}.
\ee
The third equation is simply obtained by imposing the constraint in differential form
\be
\frac{db}{dN}= \frac{d}{dN}(\frac{d\ln a_2}{d\ln a_1}).
\ee
We have to choose the value of the
 dark energy component which is determined by the parameter
\be
\frac{8\Lambda^4}{m_{\rm Pl}^2H_0^2} = c \frac{(\beta_1 +\beta_2 X_d)^2}{(m^{1111} + 3 m^{2111} X_d + 3 m^{2211} X_d^2 + m^{1222} X_d^3)}(1-\frac{\beta_1}{\beta_1^2+\beta_2^2} \Omega_m (\beta_1 +\beta_2 X_d))
\label{ci}
\ee
where for $c=1$, the dark energy component is equal to the asymptotic cosmological constant of the de Sitter space-time determined by $b=1, \ a_2 =X_d a_1$. In practice, the Universe is not in its asymptotic de Sitter phase and the coefficient $c={\cal O}(1)$ is chosen to match the $75\%$ of dark energy now.
This is achieved using the  effective dark energy fraction defined by
\be
\Omega_{\rm DE}= \bar H_J^2 - \frac{\Omega_m^{(0)}}{a_J^3}
\ee
which must be around $75\%$ now, implying a tuning of the $c$ parameter.
The effective equation of state of dark energy is obtained using
\be
3(1+\omega_{\rm DE})= \frac{d\ln \Omega_{\rm DE}}{d\ln a_J}
\ee
which must be close to -1 now. Finally, we can test the evolution of Newton's constant by calculating $\frac{d\ln \mu}{d\ln a_J}$ and comparing it to the bound (\ref{Gdot}) at the $0.02$ level by the Lunar Ranging experiment constraint \cite{Williams:2004qba}.

\subsection{Numerical results}
In the previous sections, we have described solution the different cosmological eras where $a_2=Xa_1$ and  $X$ are constant. Numerically, we will veer away from this case and explore what happens when initially $b_{\rm ini} = 1$ and $a_{2\rm ini}=X_m a_{\rm ini}$ at matter-radiation equality, i.e. far in the past the solution coincides with the one in the matter and radiation eras. The results  in figure 1 show the evolution of $a_2/a_1$ as a function of the Jordan frame redshift for two models defined below.  We have normalised the constant $c$ which dictates the numerical value of the graviton mass to be such that there is $75\%$ dark energy now. We find that the Hubble rate in the Jordan frame differs from its $\Lambda$-CDM counterpart by a few percent in the recent past of the Universe when the parameters of the model, i.e. the two couplings $\beta_{1,2}$ and the parameters $m^{ijkl}$ vary (see figure 2).

More precisely, we choose to analyse the evolution of the universe from matter-radiation equality $a_{\rm ini}=10^{-4}$ where we have $a_{2\rm ini}= X_m a_{1\rm ini} $ initially and $a_{1 \rm ini}= 10^{-4}/({\beta_1 +\beta_2  X_m})$. We take
$\Omega_m^{(0)}=0.25$. The initial value of $b$ is chosen to be $b=1$ and the Universe is on the
matter dominated explicit solution.

\subsubsection{Model I} \label{subsecI}

We consider a model where $\beta_1=2,\ \beta_2=1$ and all the $m^{ijkl}=1$. This implies that $X_d=1$ and $X_m=0.5$.
We find that $b$ varies significantly only when dark energy becomes important before  converging to its asymptotic value $b=1$ in the dark energy era (figure 3).  We can always adjust the constant $c\sim 0.715$  to  obtain around $75\%$ dark energy with an equation of state around -1 (figure 4). The background cosmology differs from $\Lambda$-CDM in the recent past (figure 2).   Moreover we have $\eta\ne 1$, $\mu\ne 1$ and $\Sigma\ne 1$ (figures 5, 6 and 7). They deviate from $\Lambda$-CDM at the $10\%$ level or below.  We also find that Newton's constant varies, but less than the present bound from the Lunar Ranging experiment (figure 8).

\subsubsection{Model II} \label{subsecII}
We consider a model where $\beta_1=1,\ \beta_2=1$ and all the $m^{ijkl}=1$ apart from $m^{1111}=2$. This implies that $X_m=1$ and $X_m=0.87$.
We find that $b$ varies significantly only when dark energy becomes important, before  converging to its asymptotic value $b=1$ in the dark energy era (figure 3).  We can always adjust the constant $c\sim 1.137$  to  obtain around $75\%$ dark energy with an equation of state around -1 (figure4). The background cosmology differs from $\Lambda$-CDM in the recent past.   Moreover we have $\eta\ne 1$, $\mu\ne 1$ and $\Sigma\ne 1$. They  deviate from $\Lambda$-CDM at the $10\%$ level or below (figure 5, 6 and 7).  We also find that Newton's constant varies, but less than the present bound from the Lunar Ranging experiment (figure 8).

\subsection{Discussion}

The cosmological evolution depends on the parameters of the model. Exploring the full parameter space of the model is beyond the scope of the present paper. Here we have concentrated on models where $X_m\ne X_d$ in order to see a variation of both $X$ and $b$. We have focussed on model I, where one coupling $\beta_1$ is larger, and on another model II, where $m^{1111}$ is also enhanced. Both models show a large deviation of $b$ from one, although of different signs. This is also the case for the variation of the Hubble rate compared to $\Lambda$-CDM although the difference is less significant. Finally, we observe that large deviations in the growth of structure and lensing can also be expected. The classification and the phenomenology of these models is left for future work.

\begin{figure*}
\centering
\includegraphics[width=0.49\linewidth]{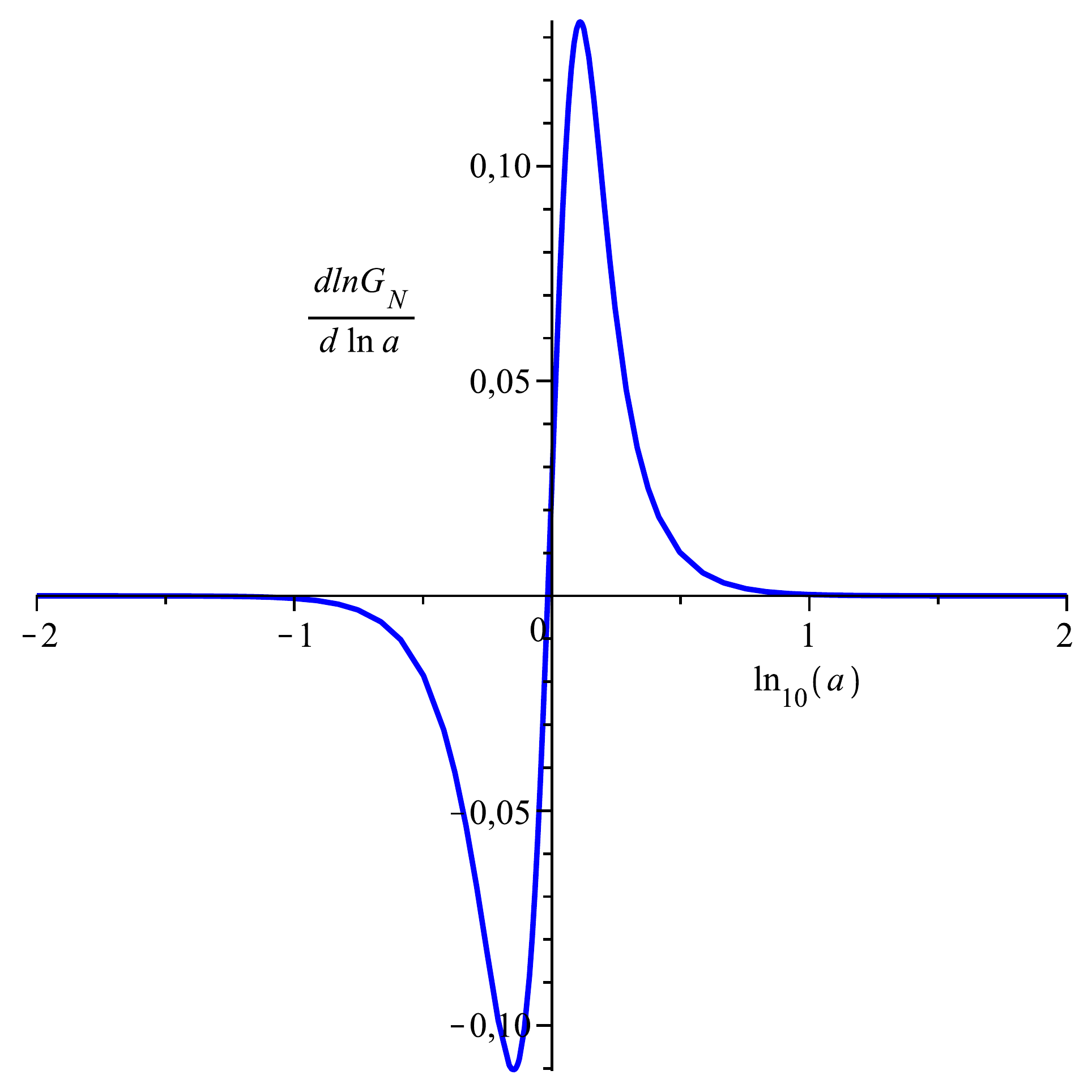}
\includegraphics[width=0.49\linewidth]{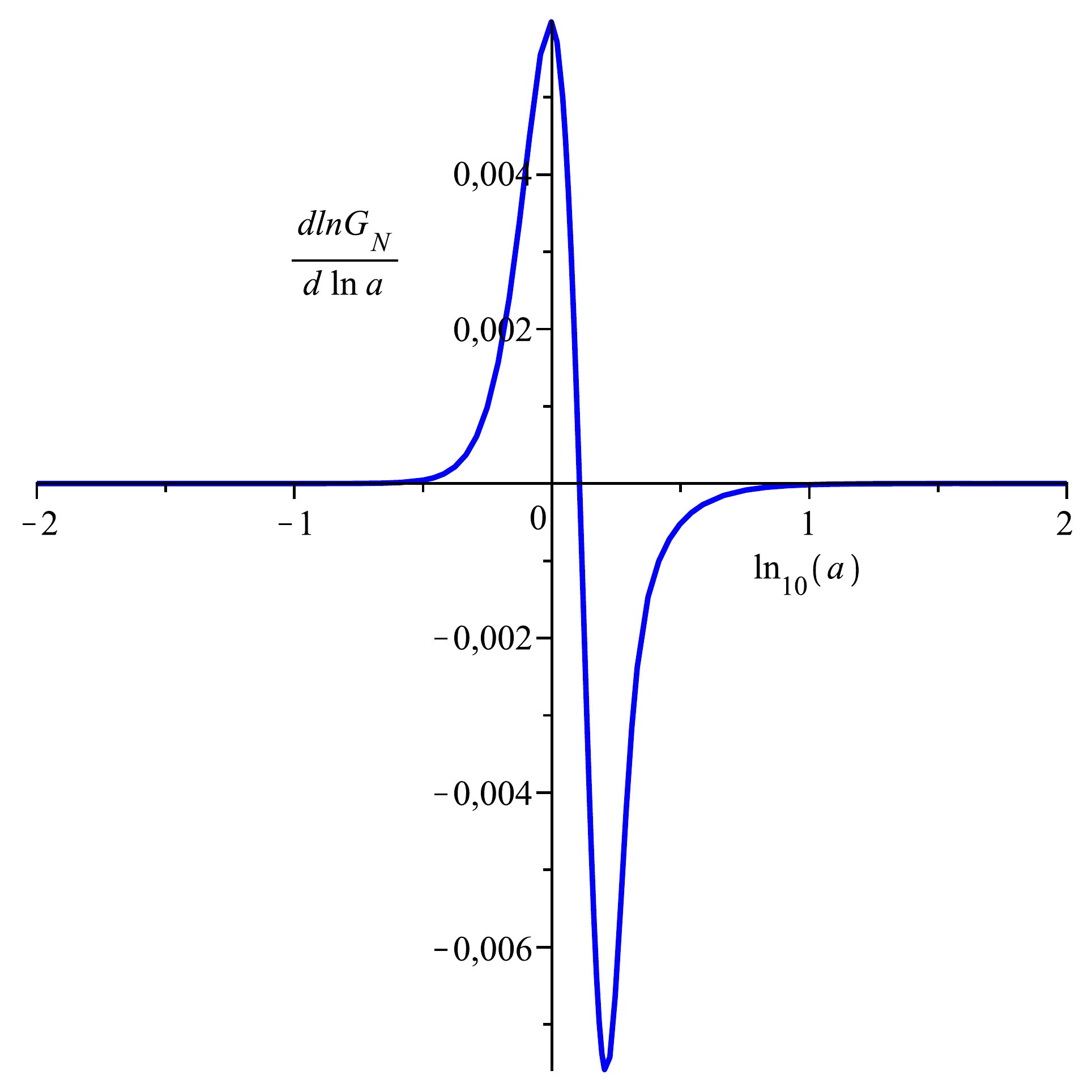}
\caption{The variation of  $\frac{d \ln G_N}{dH_J}$  as a function of  redshift $a_J$ from  $a_{\rm ini}=10^{-4}$ with model I (left panel) and model II (right panel). The bound (\ref{Gdot}) at the 0.02  level is satisfied in both cases \cite{Williams:2004qba}.}
\end{figure*}

\section{Conclusions}
We have analysed massive bigravity with a consistent matter coupling to both metrics \cite{deRham:2014naa,Noller:2014sta} in the {constrained} vielbein formalism {(equivalent to the metric formulation)}. The constrained vielbein formalism allows us to extend known properties of the metric formulation in a transparent fashion. The new results obtained in this work are as follows:
{At the background cosmological level, we have retrieved the existence of two branches of solutions for the background cosmology \cite{Enander:2014xga,Lagos:2015sya,Comelli:2015pua, Gumrukcuoglu:2015nua}. We have explicitly shown that in the asymptotic past (matter or radiation eras) and the asymptotic future (dark energy era), the ratio between the two scale factors converges to a constant and the ratio between the two lapse functions $b$ converges to unity. Deviations from these regimes only occur at the present epoch where $b$ differs from one when the degeneracy between the couplings to matter or between the coefficients of the potential term of bigravity is lifted. We have explicitly illustrated this numerically but choosing two typical examples: one where all the potential terms are on equal footing and the two matter couplings differ, and another one where the matter couplings coincide and only one of the coefficients of the potential is different from the others. We expect that more complex cases will not change drastically from the behaviour of these models. A more thorough analysis is left for future work.}

We have shown how in the quasi-static approximation, i.e. a situation which is valid in the matter era,  the scalar perturbations reduce to four Newtonian potentials. The Jordan matter and lensing properties of the model are affected by the two Newtonian potentials in the Jordan frame, which explicitly differ when the lapse functions of the two metrics differ, i.e. when $b\ne 1$. This happens only between the end of the matter era and the asymptotic future dark energy epoch. This entails that the slip parameter $\eta$, the growth parameter $\mu$ and the lensing parameter $\Sigma$ deviate from one in the recent past of the Universe, i.e. growth of structure is modified. We have also illustrated  this  explicitly  by solving the equations of motion numerically in the two sample cases described above.

{ We have examined the gravitational properties in the static case around compact objects on scales larger than the inverse cut-off and shown that GR is retrieved in this limit. This allows us to identify the local gravitational constant and identify it with the cosmological one.}

{ We have also re-examined and discussed the linear cosmological  perturbations for these theories. We have considered the instabilities of the model and given the general expression for both the graviton mass matrix and the vector mode kinetic mixing matrix in a simple and transparent way, showing that they are proportional for all models in doubly coupled bigravity. This allows us to retrieve straightforwardly that vectors and tensors suffer from instabilities in the early radiation epoch. Then and  focussing on late-time properties, i.e. in the very late radiation and  matter eras and the present epoch,  and motivated by the fact that the low-energy regime at late times offers the most robust predictions in theories with a low strong coupling scale, we have ignored the potential instabilities in the perturbative sectors (vectors and tensors) in the early Universe, already partially explored by \cite{Comelli:2015pua, Gumrukcuoglu:2015nua}. On the contrary we have only been interested in the late time regime with initial conditions set at the onset of the matter dominated era. In this case there is no vector instability, growth of structure is affected by the non-trivial parameters $(\mu,\eta,\Sigma)$  and the two tensor modes mix leading to gravitational birefringence. The study of the latter is left for future work.}
\\

\noindent {\bf Acknowledgements:} We would like to thank Emir Gumrukcuoglu, Marco Crisostomi and Luigi Pilo for useful discussions and correspondence. We would particularly like to thank Kazuya Koyama for discussions and suggestions on the vector instability part. This project has received funding from the European Union’s Horizon 2020 research and innovation programme under the Marie Skłodowska-Curie grant agreement No 690575, ACD acknowledges partial support from STFC under grants ST/L000385/1 and ST/L000636/1, JN acknowledges support from the Royal Commission for the Exhibition of 1851, BIPAC and Queen's College, Oxford.

\appendix

\section{Perturbations}

In this appendix, we present  details about the perturbative degrees of freedom of the theory. We work in the {constrained} vielbein formalism, {explicitly using} the symmetric condition (\ref{sym}) for the equivalence with the  metric formulation in the absence of matter. When matter is present, we shall use the degrees of freedom found in what follows and couple them to matter.
The scalar-vector-tensor decomposition of the linear perturbations gives {
\begin{align}
\delta e^{0\alpha} _0 &= a_\alpha \Phi_\alpha,  &\delta e^{i\alpha }_j &= -a_\alpha \Psi_\alpha \delta^i_j + a_\alpha \partial^i \partial_j U_\alpha + a_\alpha \partial_j V^i_\alpha +a_\alpha \partial^i W_{j\alpha} + a_\alpha h^i_{j\alpha} \\
\delta e^{i\alpha}_0 &= -a_\alpha\partial^i W_\alpha + a_\alpha D^i_\alpha,  &\delta e^{0\alpha}_i &= -a_\alpha\partial_i V_\alpha +a_\alpha C_{i\alpha},
\end{align}
 where the spatial index of the spatial derivative is raised with $\delta^{ij}$, i.e. $\partial^i=\delta^{ij} \partial_j$ and the index $\alpha=1,2$.  The transversality conditions are
\be
\partial^i C_{i\alpha}=0, \ \partial_i D^i_\alpha=0,\ \partial_i V^i_{\alpha}=0,\ \partial ^i W_{i\alpha}=0,\ \ \partial^i h^j_{i\alpha}=0
\ee
and tracelessness corresponds to}
\be
h^{i}_{i\alpha}=0.
\ee
The metric variation
\be
\delta g^\alpha_{i0}= a_\alpha(- b_\alpha \delta e^{0\alpha}_i+ \delta e^{i\alpha}_0)
\ee
where $b_1=1,b_2=b$, involves the combinations $b_\alpha V_\alpha- W_\alpha$ and $b_\alpha C_i^\alpha- D_i^\alpha$. We can always choose one of the two sets of perturbations to be spurious. We choose $W_\alpha=0$ and $D^i_\alpha=0$.
Similarly the metrics $g^\alpha_{ij}$ only involve the symmetric combinations $h^i_{j\alpha}+ h^{j}_{i\alpha}$ { and $V_{i\alpha}+W_{i\alpha}$}.  We set the antisymmetric parts to 0 
and therefore $h^i_{j\alpha}=h^{j}_{i\alpha}$. { We also choose $W_{i\alpha}= V_{i\alpha}$.
The symmetric condition $(0i)$ implies that
\be
V_2=bV_1, \ C_{i2}= b C_{i1}.
\ee
The $(ij)$ constraint  { is automatically satisfied.}
We have now the possibility of using four gauge transformations (\ref{gauge}) $\xi^\mu=(\xi^0,\xi^i= \tilde \xi^i +\partial^i \Theta)$ where $\partial_i \tilde \xi^i=0$.
As explicitly proved in the main text, taking $\xi^0=-V^1$ and $\Theta=-U_2$, one can gauge away $V_1$ and $U_2$, leaving only $U=U_1-U_2$ as a scalar on top of the four Newtonian potentials.
Finally taking $\tilde \xi^i=V^i_2$, one can gauge away $V^i_2$ { leaving only $V_i= V_{i1}-V_{i2}$ as a vector perturbation}. Notice that this step involves the quasi-static approximation as we neglect terms like $\partial_0 V^i_2$ which would otherwise reappear in $\delta e^i_{0 2}$. } After this gauge fixing, we are thus left with the perturbations
\be
\delta e^{0\alpha} _0= a_\alpha \Phi_\alpha, \ \delta e^{i\alpha }_j = -a_\alpha \Psi_\alpha \delta^i_j + a_\alpha \partial^i \partial_j U \delta_{\alpha 1}+ \a_{\alpha} \partial_{\{ j}V^{i\}}\delta_{\alpha 1} + a_\alpha h^i_{j\alpha}
\ee
and
\be
 \delta e^{0\alpha}_i= a_\alpha C_{i\alpha}.
\ee
As a result, the perturbations comprise the four Newtonian potentials $(\Phi_\alpha,\Psi_\alpha)$, the scalar $U$, the two vectors $(C_{i1}, V_i)$ and the gravitons $h^i_{j\alpha}$. In a Minkowski
background where all the Newton potentials vanish, this reduces to seven degrees of freedom as expected for a massive graviton and one massless one, { together with one divergenceless vector $V_i$ which decouples from pressure-less matter}.

{ Let us consider now what happens if a different gauge choice is made {and one goes beyond the quasi-static approximation}. The symmetric condition prior to any gauge choice is such that its (0i) part leads to
\be
W_1+ bV_1= W_2+V_2,\  D_1^i+b C^i_1=D_2^i + C^i_2 \label{sol}
\ee
{ and its (ij) part to
\be
V^i_1-W^i_1=V^i_2-W^i_2.
\ee
It is convenient to define the two symmetric combinations
\be
S^i_\alpha=\frac{1}{2}(V^i_\alpha+ W^i_\alpha)
\ee
and the antisymmetric one
\be
A^i=\frac{1}{2}(V^i_1- W^i_1)=\frac{1}{2}(V^i_2- W^i_2).
\ee
Under a gauge transformation $(0, \tilde \xi^i +\partial^i \Theta)$, we have that
\be
V^i_\alpha\to V^i_\alpha -\frac{\tilde \xi^i}{2},\ W^i_\alpha\to W^i_\alpha -\frac{\tilde \xi^i}{2}, \ D^i_\alpha\to D^i_\alpha -\partial_0 \tilde \xi^i
\ee
and
\be
V_\alpha \to V_\alpha, \ W_\alpha \to W_\alpha +\partial_0\Theta.
\ee}
The metrics $\delta g^\alpha_{i0}$ only involve the combinations $b_\alpha V_\alpha- W_\alpha$ and $ b_\alpha C_i^\alpha-D_i^\alpha$. As a result the physics only depends on two out of the four fields  $(C^i_\alpha,D^i_\alpha)$ and $(V_\alpha,W_\alpha)$ respectively. Hence one can choose linear gauges which are linearly independent of $b_\alpha V_\alpha+ W_\alpha$ and $ b_\alpha C_i^\alpha+ D_i^\alpha$
\be
G_\alpha=c_\alpha V_\alpha+ d_\alpha W_\alpha\equiv 0
\ee
and
\be
G^i_\alpha= \tilde c_\alpha C_i^\alpha+ \tilde d_\alpha D_i^\alpha\equiv \vec 0
\ee
i.e. such that $c_\alpha/d_\alpha + b_\alpha\ne 0$ and  $\tilde c_\alpha/\tilde d_\alpha + b_\alpha\ne 0$. This allows one to express $W_\alpha$ as a function of $V_\alpha$, and $D^i_\alpha$ as a function of $C^i_\alpha$
This reduces the four
variables $(C^i_\alpha,D^i_\alpha)$ and $(V_\alpha,W_\alpha)$ respectively to one vector and one scalar.
{ Similarly $\delta g^{\alpha}_{ij}$ only depends $S_{\alpha i}$. This allows one to set
\be
A^i\equiv 0.
\ee
These gauge choices transform as
\be
G_\alpha \to G_\alpha +d_\alpha \partial_0\Theta
\ee
and
\be
G^i_\alpha \to G^i_\alpha-\tilde d_\alpha \partial_0 \tilde \xi^i, \ A^i\to A^i
\ee
under  the diagonal copy of diffeomorphism invariance.
Under the remaining gauge transformations parameterised by $\xi^0$, we have
\be
V^i_\alpha\to V^i_\alpha,\ W^i_\alpha \to W^i_\alpha, \  D^i_\alpha\to D^i_\alpha
\ee
and
\be
V_\alpha \to V_\alpha+ b_\alpha \xi^0, \ W_\alpha \to W_\alpha.
\ee
Therefore the gauge conditions transform as
\be
G_\alpha \to G_\alpha +c_\alpha b_\alpha \xi^0
\ee
and
\be
G^i_\alpha \to G^i_\alpha,\ A^i\to A^i.
\ee
In general fixing the gauge arbitrarily leaves no residual gauge symmetry. Therefore one finds that there is one degree of freedom left amongst $(C^i_\alpha,D^i_\alpha)$ and $(V_\alpha,W_\alpha)$ respectively, say $C^i_1$ and $V_1$. The two vectors $V^i_\alpha=W^i_\alpha$ are  also present. We are thus left with seven scalars $(\Psi_\alpha,\Phi_\alpha,U_\alpha, V_1)$, three vectors, $C^i_1$ and $V^i_\alpha=W^i_\alpha$,  and two tensors.}

This is not a clever choice as two different gauge choices allow one to reduce the number of degrees of freedom further. The first one corresponds to  $d_\alpha=\tilde d_\alpha=0$ which preserves gauge invariance parameterised by $(\Theta, \tilde \xi^i)$ and breaks the one given by $\xi^0$. Choosing $\partial_0 \tilde \xi^i= D^i_1=D^i_2$ and $\partial_0 \Theta= -W_1=-W_2$ now that $V_1=V_2=0$ and $C^i_1=C^i_2=0$, we find that all the fields $(C^i_\alpha,D^i_\alpha)$ and $(V_\alpha,W_\alpha)$ are projected away. { We are thus left with six scalars $(\Psi_\alpha,\Phi_\alpha,U_\alpha)$, two vectors $V^i_\alpha$ and two tensors.
Another choice corresponds to $c_\alpha=\tilde c_\alpha=0$ which breaks the gauge invariance parameterised by $(\Theta, \tilde \xi^i)$ and preserves  the one given by $\xi^0$. Choosing $\xi^0=-V_1$ allows one to remove one scalar. The remaining fields are the six scalars $(\Psi_\alpha,\Phi_\alpha,U_\alpha)$, three vectors $C^i_1$ and $V^i_\alpha$ and two tensors.}

 In the quasi-static approximation where $\partial_0 \Theta\sim 0$ and $\partial_0 \tilde \xi^i\sim 0$ and choosing the gauge $c_\alpha=\tilde c_\alpha=0$, { we retrieve the gauge freedom parameterised by $(\Theta, \tilde \xi^i)$ which allows one to reduce the number of degrees of freedom, in particular one can gauge away the  vector $V^i_2$ and the extra scalar $U_2$. Hence in the quasi-static approximation the minimal number of degrees of freedom comprises five scalars $(\Psi_\alpha,\Phi_\alpha,U)$, two vectors $C^i_1$ and $V^i$,  and two tensors. } This shows that the number of degrees of freedom and their dynamics simplify drastically in the quasi-static approximation. In particular this demonstrates that the quasi-static approximation allows one to remove one scalar degree of freedom.
{

In conclusion, we find that in a general time-dependent situation, by choosing the gauge condition where $d_\alpha=\tilde d_\alpha=0$, the spectrum of cosmological perturbations reduces to six scalars $(\Psi_\alpha,\Phi_\alpha,U_\alpha)$, two vectors $V^i_\alpha$ and two tensors. In a Minkowski background with static sources, such as stars with small Newtonian potentials, the fields are static with no time dependence. In this case, the quasi-static results apply and one can reduce the number of degrees of freedom to five scalars $(\Psi_\alpha,\Phi_\alpha,U)$, two vectors $C^i_1$  and $V^i$ and two tensors. In the absence of external static sources the Newtonian potentials
vanish, $V^i$ decouples from pressure-less matter and one is {manifestly} left with two gravitons, one massless and another massive one as expected in bigravity.}

\bibliographystyle{JHEP}
\bibliography{bgde}

\end{document}